\documentclass[twocolumn,twocolappendix]{aastex631}

\newcommand{\LzLc}{$L_\mathrm{z}/L_\mathrm{c}$}

\def\vrot{v_{\rm rot}}

\def\SFR{\dot{M}_{\star}}

\def\Msunyr{{\rm \;M_\odot\;yr^{-1}}}
\def\Msun{{\rm \;M_\odot}}
\def\Myr{{\rm \;Myr}}

\def\afe{{[$\alpha$/Fe]}}
\def\feh{{[Fe/H]}}
\newcounter{sum}
\usepackage{CJKutf8}

\defcitealias{Semenov2023}{S23}

\shorttitle{The Three-Phase Evolution of the Milky Way}
\shortauthors{Chandra et al.}

\graphicspath{{./}}

\begin{document}

\title{The Three-Phase Evolution of the Milky Way}

\author[0000-0002-0572-8012]{Vedant Chandra}
\affiliation{Center for Astrophysics $\mid$ Harvard \& Smithsonian, 60 Garden St, Cambridge, MA 02138, USA}
\affiliation{Max-Planck-Institut f{\"u}r Astronomie, K{\"o}nigstuhl 17, D-69117 Heidelberg, Germany}

\correspondingauthor{Vedant Chandra}
\email{vedant.chandra@cfa.harvard.edu}

\author[0000-0002-6648-7136]{Vadim A. Semenov}
\affiliation{Center for Astrophysics $\mid$ Harvard \& Smithsonian, 60 Garden St, Cambridge, MA 02138, USA}

\author[0000-0003-4996-9069]{Hans-Walter Rix}
\affiliation{Max-Planck-Institut f{\"u}r Astronomie, K{\"o}nigstuhl 17, D-69117 Heidelberg, Germany}

\author[0000-0002-1590-8551]{Charlie Conroy}
\affiliation{Center for Astrophysics $\mid$ Harvard \& Smithsonian, 60 Garden St, Cambridge, MA 02138, USA}

\author[0000-0002-7846-9787]{Ana~Bonaca}
\affiliation{The Observatories of the Carnegie Institution for Science, 813 Santa Barbara Street, Pasadena, CA 91101, USA}

\author[0000-0003-3997-5705]{Rohan~P.~Naidu}
\altaffiliation{NASA Hubble Fellow}
\affiliation{MIT Kavli Institute for Astrophysics and Space Research, 77 Massachusetts Ave., Cambridge, MA 02139, USA}

\author[0000-0001-8006-6365]{Ren\'e Andrae}
\affiliation{Max-Planck-Institut f{\"u}r Astronomie, K{\"o}nigstuhl 17, D-69117 Heidelberg, Germany}

\author[0000-0002-3651-5482]{Jiadong Li \begin{CJK*}{UTF8}{gbsn}(李佳东)\end{CJK*}}
\affiliation{Max-Planck-Institut f{\"u}r Astronomie, K{\"o}nigstuhl 17, D-69117 Heidelberg, Germany}

\author[0000-0001-6950-1629]{Lars Hernquist}
\affiliation{Center for Astrophysics $\mid$ Harvard \& Smithsonian, 60 Garden St, Cambridge, MA 02138, USA}

\begin{abstract}

\noindent
We illustrate the formation and evolution of the Milky Way over cosmic time, utilizing a sample of 10~million red giant stars with full chemodynamical information, including metallicities and $\alpha$-abundances from low-resolution \textit{Gaia} XP spectra.
The evolution of angular momentum as a function of metallicity --- a rough proxy for stellar age, particularly for high-[$\alpha$/Fe] stars --- displays three distinct phases: the disordered and chaotic protogalaxy, the kinematically-hot old disk, and the kinematically-cold young disk. 
The old high-$\alpha$ disk starts at [Fe/H]~$\approx -1.0$, `spinning up' from the nascent protogalaxy, and then exhibits a smooth `cooldown' toward more ordered and circular orbits at higher metallicities. 
The young low-$\alpha$ disk is kinematically cold throughout its metallicity range, with its observed properties modulated by a strong radial gradient.
We interpret these trends using Milky Way analogs from the TNG50 cosmological simulation, identifying one that closely matches the kinematic evolution of our Galaxy. 
This halo's protogalaxy spins up into a relatively thin and misaligned high-$\alpha$ disk at early times, which is subsequently heated and torqued by a major gas-rich merger. 
The merger contributes a large amount of low-metallicity gas and angular momentum, from which the kinematically cold low-$\alpha$ stellar disk is subsequently born. 
This simulated history parallels several observed features of the Milky Way, particularly the decisive `GSE' merger that likely occurred at $z \approx 2$. 
Our results provide an all-sky perspective on the emerging picture of our Galaxy's three-phase formation, impelled by the three physical mechanisms of spinup, merger, and cooldown.

\end{abstract}


\section{Introduction} \label{sec:intro}

It is a central goal of modern astronomy to piece together a coherent formation history of the Milky Way (MW). More broadly, we wish to understand how much formation memory a galaxy retains over cosmic timescales, and consequently investigate the physical mechanisms that shape disk galaxies. 
For both aspects, the Milky Way is the ideal `near-field' laboratory with its detailed star-by-star information. 
Finally, the Milky Way is set apart from all other galaxies in the Universe because it hosts us, and it is inherently desirable to understand how our own cosmic home came to be.

From a theoretical perspective, disk galaxies have long been understood to emerge as a consequence of dissipative gas collapse in dark matter halos, and the retention of angular momentum acquired during large-scale structure formation \citep{Eggen1962, Mestel1963, Peebles1969, White1978, Fall1980, Mo1998}. 
These ideas formed the basis of subsequent analytic and semi-analytic models which succeeded in producing reasonably realistic disk galaxies \citep[e.g.,][]{Silk1997, Dalcanton1997, Dekel2009, Dekel2013, Somerville2015, Stevens2016}. 
Similar progress has been made in numerical simulations, with a variety of works exploring disk formation in cosmological simulations, from idealized and zoom-in scales to large volume scales \citep[e.g.][]{Bournaud2009, Hopkins2009, Hopkins2023, Minchev2012, Bird2013, Agertz2015, Genel2015, Schaye2015,  Agertz2016, Grand2017, Naab2017, Garrison-Kimmel2018, Jiang2019, Tacchella2019, Pillepich2019, Stern2021, Stern2023, Yu2021, Yu2023, Agertz2021, Dillamore2022, Hafen2022, Khoperskov2022a, Khoperskov2022c, Gurvich2023, Wang2023, Pillepich2023, Semenov2023, Semenov2023b, Santucci2023}.
These simulations showed the crucial importance of feedback-driven galactic outflows---which retain the angular momentum that galaxies inherit from their dark matter halos---as well as the importance of stochastic and occasionally violent processes like mergers and variable gas accretion flows.

The physical mechanism of disk formation is still a subject of active debate. For example,
\citet[][see also \citealt{Hafen2022}]{Stern2021, Stern2023} put forward a model where cold disk formation is triggered by the assembly of a hot inner circumgalactic halo, and an associated reduction in the stochasticity of star formation. 
On the other hand, \cite{Hopkins2023} use controlled numerical experiments based on FIRE-2 cosmological simulations to argue that the fundamental \textit{causative} variable governing disk formation is a centrally concentrated density profile, which suppresses destructive radial gas motions and permits the coherent growth of angular momentum. 

Recently, \citet[][hereafter \citetalias{Semenov2023}]{Semenov2023} exploited the diversity of MW-like galaxies in the Illustris TNG50 cosmological simulations \citep{Pillepich2018, Nelson2019} to interpret observational evidence that the MW disk formed atypically early \citep{Belokurov2022a,Belokurov2023,Conroy2022}.
They corroborate the idea that disk formation is most closely linked with mass assembly, suggesting that the MW may have experienced unusually rapid mass growth at early times. 
At the same time, the underlying physical driver of disk formation is hard to pinpoint as both hot halo formation and gravitational potential steepening tend to occur concurrently with the disk spinup \citep{Semenov2023b}, and the potential steepening continues throughout the subsequent evolution of the disk.
This suggests that a centrally concentrated potential might be a consequence of disk formation rather than its cause \citep{Dillamore2023}. 

In the past decade, our perspective of the Milky Way has expanded greatly through the proliferation of photometric, chemical, and kinematic data for millions of stars \citep[see the reviews of][]{Ivezic2012,Rix2013,Bland-Hawthorn2016}. 
It is reasonably certain that we reside in a barred spiral galaxy, with two key characteristics of the MW being a total stellar mass $\approx 5 \times 10^{10}\,M_\odot$ and a relatively quiescent present-day star formation rate of $\approx 1.6\,M_\odot\,\mathrm{yr}^{-1}$ \citep{Licquia2015, Bland-Hawthorn2016}. 
The spatial, kinematic, chemical, and temporal structure of the disk has been exquisitely mapped since the advent of all-sky spectroscopic surveys, augmented in the past half-decade with astrometry from the \textit{Gaia} space observatory \citep[e.g.,][]{Bovy2012a, Bovy2012b, Haywood2013, Bensby2014, Hayden2015, Bovy2016a, Weinberg2019, Weinberg2022, Xiang2022, Drimmel2023, RecioBlanco2023}. 

Apart from a potentially early-forming disk, there are several other features that set the MW apart from typical disk galaxies. 
From an integrated perspective, the MW lies in the intermediate `green valley' between most star forming and quiescent galaxies \citep{Mutch2011, Licquia2015}. 
Furthermore, it is exceptionally disk-dominated compared to other known galaxies with comparable stellar masses \citep{Kent1991, Kauffmann2003}. 
Finally, the MW is not isolated, and belongs to a `Local Group' of galaxies with a nearby pair of star-forming satellites, the Magellanic Clouds. 

The MW disk is known to have a striking bimodality in the abundance of $\alpha$ elements at constant metallicity \citep[e.g.,][]{Fuhrmann1998, Bensby2014, Hayden2015}. 
One way to explain this is through the infall of fresh gas from a major merger, which would dilute the overall metallicity of the star-forming gas without significantly changing the abundance ratios \citep{Chiappini1997, Grisoni2017, Buck2020, Buck2020a, Renaud2021a, Renaud2021b}. 
However, it is also possible to produce this chemical bimodality solely through secular processes of chemical enrichment and radial migration \citep{Schonrich2009a, Schonrich2009b, Minchev2013, Loebman2016, Sharma2021, Johnson2021}. 
There is tentative evidence from \textit{JWST} that such a chemical bimodality is absent from our neighbor, Andromeda \citep{Nidever2023}. 
 
Recent work has advanced our understanding of various components of the Galaxy, including the ancient protogalaxy \citep[e.g.,][]{Rix2022, Horta2023}, the birth of the disks \citep[e.g.,][]{Xiang2022}, the in-situ halo \citep[e.g.,][]{Bonaca2020, Belokurov2020b}, and the accreted halo \citep[e.g.,][]{Ibata1994, Helmi1999, Majewski2003, Newberg2009, Nissen2010, Belokurov2018a, Helmi2018, Naidu2020, Bonaca2020b, Malhan2022a}. 
Most recently, \cite{Belokurov2022a} isolate \textit{in-situ} stars in the APOGEE survey and find a dramatic increase in the rotational support of stars at \feh{}~$\approx -1.0$, marking the transition between the protogalaxy (`Aurora' in their terminology) and the ancient disk (see also \citealt{Rix2022}).
\cite{Conroy2022} study this transition with abundances and ages from the H3 Survey, finding that it coincides with a nonmonotonic rise in \afe{} abundances, suggesting a near-instantaneous transition in the star formation efficiency or gas inflow \citep[see][]{Chen2023}.
Here, we seek to provide a global perspective on these results using the latest data from the \textit{Gaia} space observatory. 

The paper is structured as follows.
We summarize the construction of our data set --- 10 million stars with full chemodynamical information from \textit{Gaia} DR3 --- in $\S$\ref{sec:data}. 
In $\S$\ref{sec:birth}, we present a unified perspective on the formation of the Milky Way, focusing on the observed three-phase evolution of angular momentum over cosmic time. 
Our goal is to present a compact and cohesive set of figures that qualitatively summarize the evolution of the Galaxy, and provide clues about its history. 
Towards this aim, in $\S$\ref{sec:tng} we interpret our data using simulations of MW analogs from the IllustrisTNG50 cosmological simulations, focusing on one particularly similar analog as an illustrative example. 
We discuss and contextualize the three-phase birth of the MW in $\S$~\ref{sec:discuss}, and summarize our conclusions in $\S$\ref{sec:conclude}.

\section{Gaia DR3 Data}\label{sec:data}

The third data release of the \textit{Gaia} space observatory provided low-resolution `XP' prism spectroscopy for $\approx 200$~million stars brighter than $G = 17.65$, and radial velocities for a subset of $\approx 30$~million stars \citep{GaiaCollaboration2022,DeAngeli2022,Montegriffo2022a,Katz2022}. 
Subsequent work has augmented this dataset by producing precise estimates of the overall metallicity \feh{} and alpha-abundance \afe{} using XP spectra \citep[][]{Andrae2023, Zhang2023, Leung2023, Li2023}. 
This sample is powerful not only because of its unprecedented size and homogeneity, but also because of its all-sky coverage and simple selection function, unlike previous pencil-beam spectroscopic surveys. 
This is therefore the ideal dataset to qualitatively explore a unified picture of our Galaxy's formation. 

Our final observational sample contains 9,907,115 RGB stars with metallicities, [$\alpha$/Fe] abundances, and full 6D phase space coordinates. Here, we outline the construction of this dataset.

\begin{figure}
    \centering
    \includegraphics[width=\columnwidth]{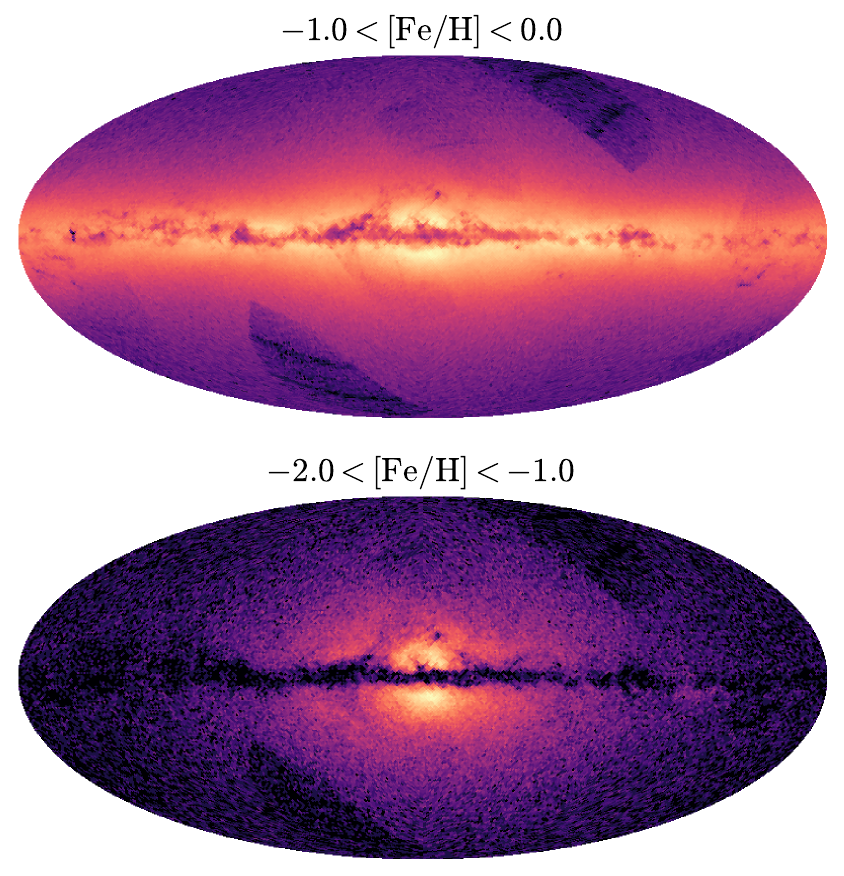}
    \includegraphics[width=0.9\columnwidth]{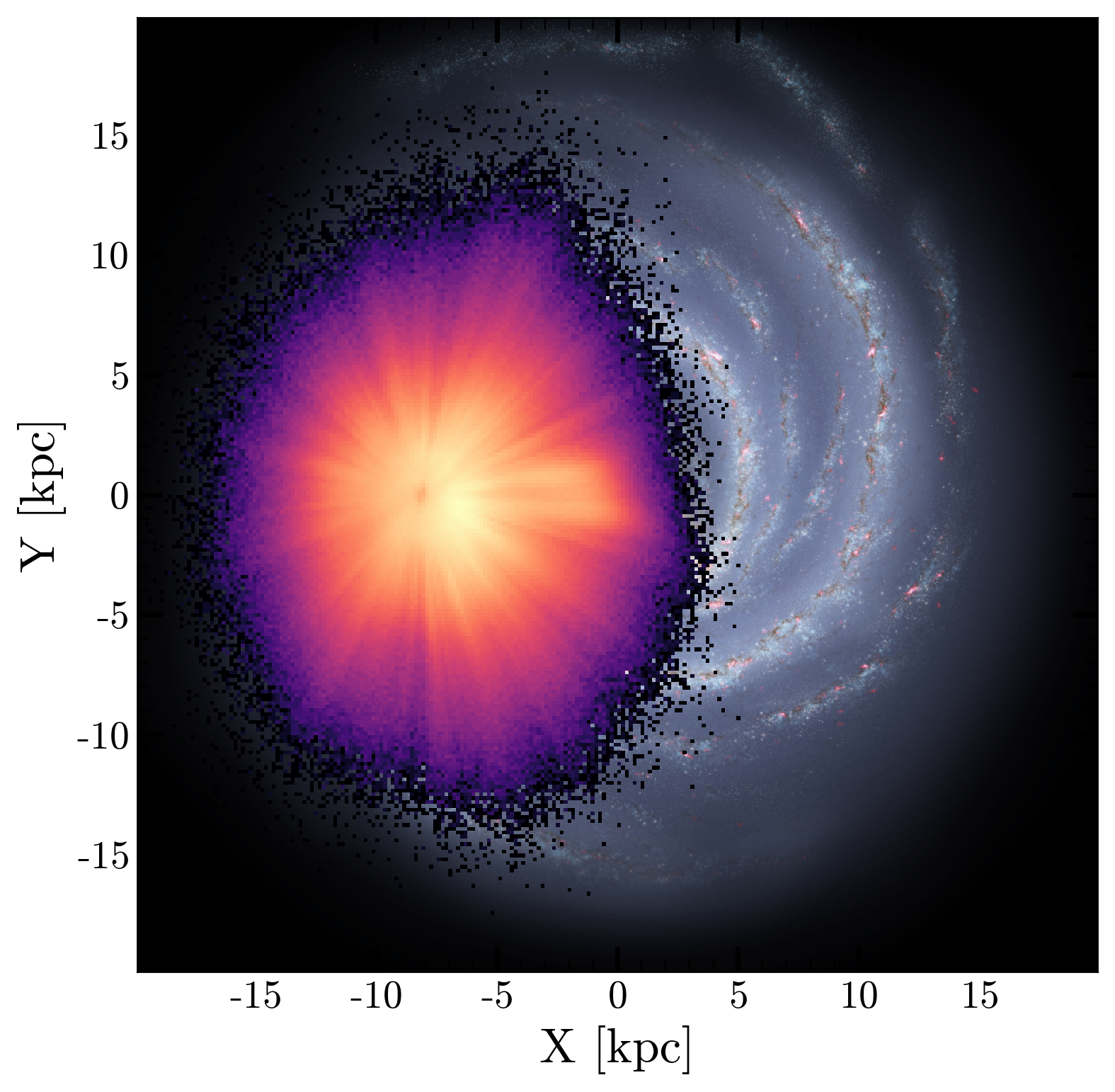}
    \caption{Top and middle: Logarithmic Galactic map of our RGB sample from \textit{Gaia} DR3, for two \textit{Gaia} XP metallicity selections. Several selection effect artifacts are visible, primarily due to \textit{Gaia}'s scanning law; however, these spatial variations are below the $\approx 10\%$ level.
    Bottom: Galactocentric distribution of our sample, overlaid on an artist's rendering of the Milky Way disk (NASA/JPL-Caltech/R. Hurt).}
    \label{fig:allsky}
\end{figure}

\subsection{Metallicities and \afe{} with \textit{Gaia} XP}

\citet{Andrae2023} construct a data-driven `XGBoost' model to estimate stellar metallicity [M/H] from low-resolution XP spectra, utilizing a training set from the Apache Point Observatory Galactic Evolution Experiment (APOGEE; \citealt{Majewski2017}), augmented with metal-poor stars from \citet{Li2022c}. 
These metallicities have been validated to be precise and accurate compared to external spectroscopic surveys. 
Although the model formally predicts bulk metallicity [M/H], the APOGEE values in the training set are calibrated to match the iron abundance [Fe/H]. 
We therefore utilize the notation [Fe/H] for metallicity throughout this work. 
For more details on feature selection and catalog validation, we refer to \cite{Andrae2023}. 

From the metallicity sample of \cite{Andrae2023}, we construct a clean sample of RGB stars with \textit{Gaia} DR3 radial velocities. 
Specifically, we adopt the following selection criteria: 

\begin{itemize}
\itemsep0em 
\item $\texttt{phot\_bp\_mean\_mag} < 16.5$
\item $\varpi/\sigma_\varpi>5$
\item $\texttt{teff\_xgboost}<5200$
\item $\texttt{logg\_xgboost}<3.5$
\item $M_{W1} > -0.3 - 0.006\cdot(5500-\texttt{teff\_xgboost})$
\item $M_{W1} > -0.01\cdot(5300-\texttt{teff\_xgboost})$
\item $(G - W_2) < 0.2 + 0.77 (G_\mathrm{BP} - W_1)$
\end{itemize}
where \texttt{teff\_xgboost} and \texttt{logg\_xgboost} are from \cite{Andrae2023}, $\varpi$ is the \textit{Gaia} parallax, and $M_{W1} = W_1 + 5\cdot \log_{10}(\varpi/100)$ is the absolute magnitude in the WISE W1 passband. 
The first two cuts select bright stars with reliable parallaxes and enough flux in the bluer BP spectrum to derive robust metallicities. 
The temperature selection removes hot stars that have spuriously low measured metallicities, and the final four cuts isolate red giants on the Hertzsprung-Russel diagram.
We excise 6841 stars that have a $> 50\%$ probability of residing in a globular cluster according to \cite{Vasiliev2021b}.
This leaves us with $\approx 11$~million RGB stars with \textit{Gaia} metallicities and radial velocities.

\citet{Li2023} present a catalog of \afe{} measurements inferred from \textit{Gaia} XP spectra through a neural network model. 
This catalog has been cross-validated to match existing spectroscopic surveys, with median RMS errors of $\approx 0.05$~dex in \afe{} for stars in our sample. 
We cross-match our clean RGB sample to this catalog, obtaining our final sample of $\approx 10$~million stars with \afe{} information. 
The spatial distribution of these stars is illustrated in Figure~\ref{fig:allsky}. 
Simply selecting metal-poor stars on this all-sky map reveals the protogalactic population of the Milky Way \citep{Rix2022}, whose evolution is further investigated in our work. 
Our sample uniformly covers a significant fraction of the Milky Way's stellar disk (bottom panel of Figure~\ref{fig:allsky}). 

\begin{figure}
    \centering
    \includegraphics[width=\columnwidth]{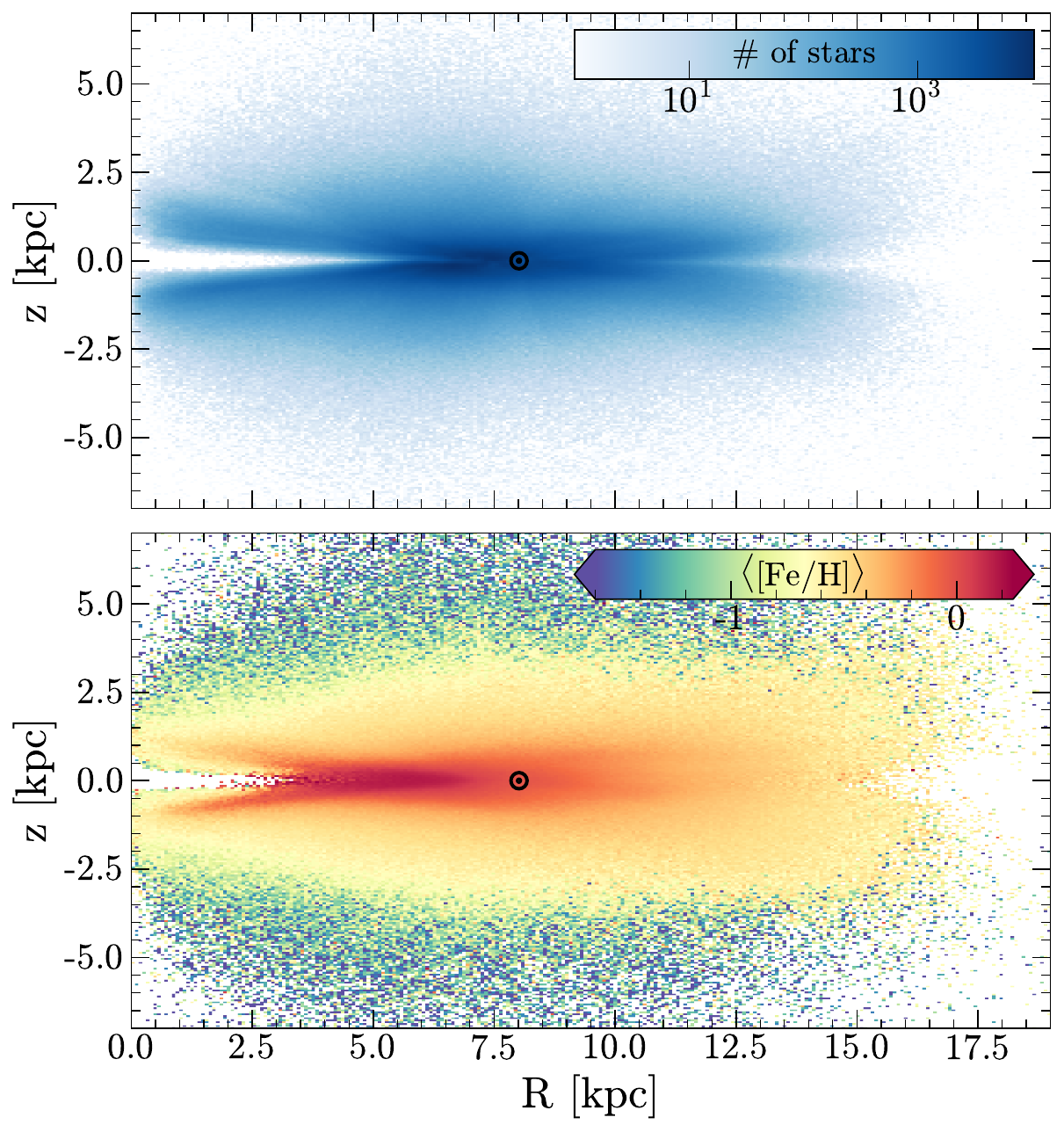}
    \caption{Summary of our \textit{Gaia} sample in cylindrical Galactocentric coordinates. Top: Logarithmic density of RGB stars with XP metallicities and \textit{Gaia} RVs in the Galactocentric $R$-$z$ plane. The survey selection is strongly mediated by dust extinction near the midplane. Bottom: The $R$-$z$ plane colored by median XP [Fe/H]. There is a strong negative metallicity gradient with increasing $R$ and $|z|$. }
    \label{fig:Rz}
\end{figure}

Figure~\ref{fig:Rz} summarizes our sample in Galactocentric radius $R$ and midplane height $z$ coordinates. In the top panel (log density of stars), two key selection effects are visible. First, most of our sample is restricted to $R \lesssim 15$ and $z \lesssim 8$~kpc, a consequence of the relative brightness of our sample with \textit{Gaia} RVs ($G < 16.5$). This is also the volume within which parallaxes can provide reliable geometric distances, satisfying our selection of $\varpi / \sigma_\varpi > 5$. Future work could extend this sample to larger distances by utilizing the full spectrophotometric information content of these stars \citep[e.g.,][]{Hogg2019}.

\begin{figure}
    \centering
    \includegraphics[width=\columnwidth]{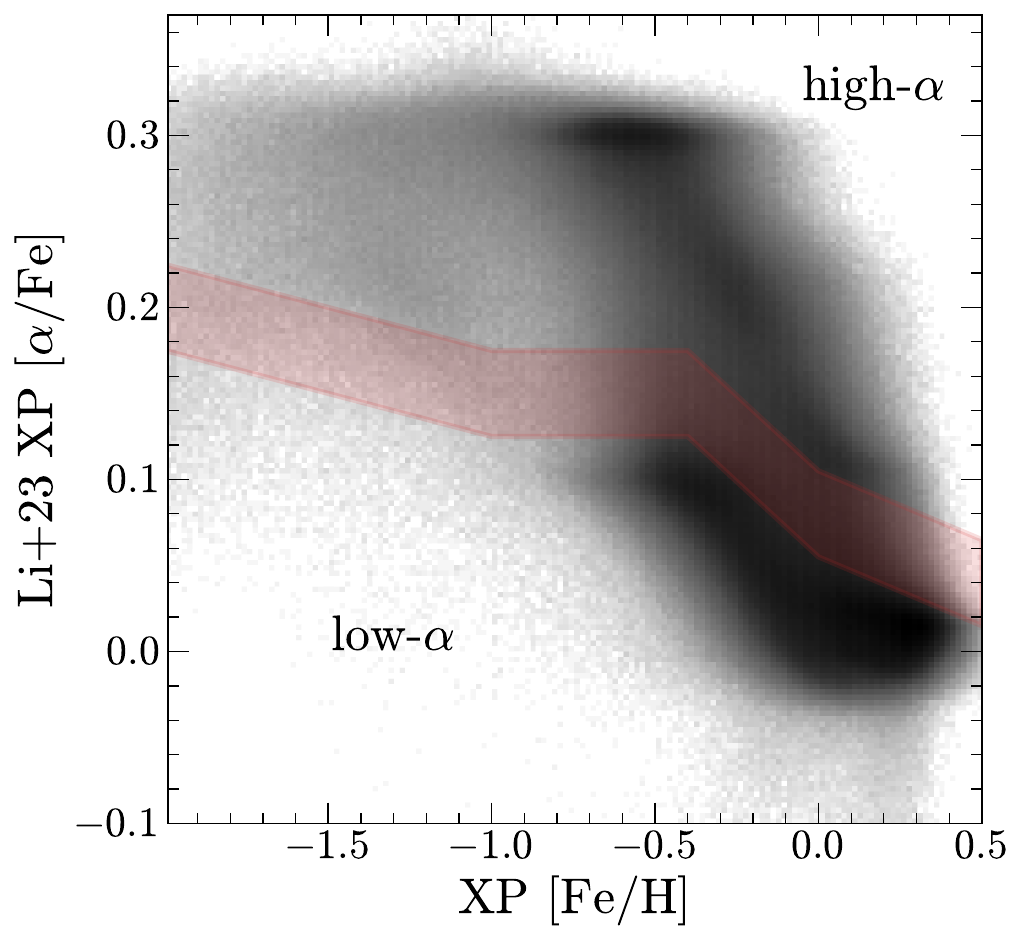}
    \caption{Logarithmic density of the all-sky \textit{Gaia} sample in abundance space, using XP-inferred [$\alpha$/Fe] abundances from \cite{Li2023}. The red band demarcates our high-$\alpha$ and low-$\alpha$ samples, with stars lying in the intermediate region being excluded to keep the samples pure.}
    \label{fig:afe}
\end{figure}

Figure~\ref{fig:afe} shows the XP-inferred stellar abundance plane, in which the two canonical disk populations --- high-$\alpha$ and low-$\alpha$ --- are visible. 
These definitions are often loosely mapped onto the spatial populations of `thick' and `thin' disks respectively. 
As will be demonstrated in this work (and has already been argued, e.g., \citealt{Rix2013, Bensby2014, Hayden2015}), the chemical bimodality maps into spatial differences, but there is no clear bimodality in the vertical structure: the high- and low-$\alpha$ disks differ more in their radial structure than their thickness. 
Therefore, we use this abundance space to differentiate the two populations, and separately study their spatial and kinematic configuration. 

The red lines in Figure~\ref{fig:afe} show the selection cuts applied to select stars from both disk populations. 
Stars in the ambiguous region between the two selections are omitted, to ensure that we have pure samples of both high-$\alpha$ and low-$\alpha$ disk stars. 
This is particularly important when studying the high-$\alpha$ disk, since even modest \afe{} errors can lead to significant contamination from the much more densely populated low-$\alpha$ disk.

\subsection{6D Phase Space Coordinates}

For each of RGB star in our sample, the Bayesian prior-informed geometric distance from \cite{Bailer-Jones2020} is adopted, which combines the \textit{Gaia} parallax with a Galactic space density prior. Our results are not significantly affected if we use zero-point-corrected inverse-parallax distances instead, but the \cite{Bailer-Jones2020} distances provide a more principled framework to infer distances across all SNR regimes. 

Armed with 6D phase-space information for these RGB stars, we compute their Galactocentric coordinates (Figure~\ref{fig:allsky}), azimuthal angular momenta $L_\mathrm{Z}$, and total energies $E$. The latter requires an estimate of the Galactic potential, for which we adopt a four-component Milky Way model containing spherical \citet{Hernquist1990} nucleus and bulge components, an (approximate) exponential disk component \citep{Smith2015}, and a spherical NFW halo component. The potential is fit to the same measurements of the enclosed mass as the \texttt{MilkyWayPotential} in \texttt{gala} \citep{gala, adrian_price_whelan_2020_4159870}, with the added constraint of having a circular velocity at the solar position $= 229~\textrm{km}~\textrm{s}^{-1}$ \citep{Eilers2019}. 

\begin{figure}
    \centering
    \includegraphics[width=\columnwidth]{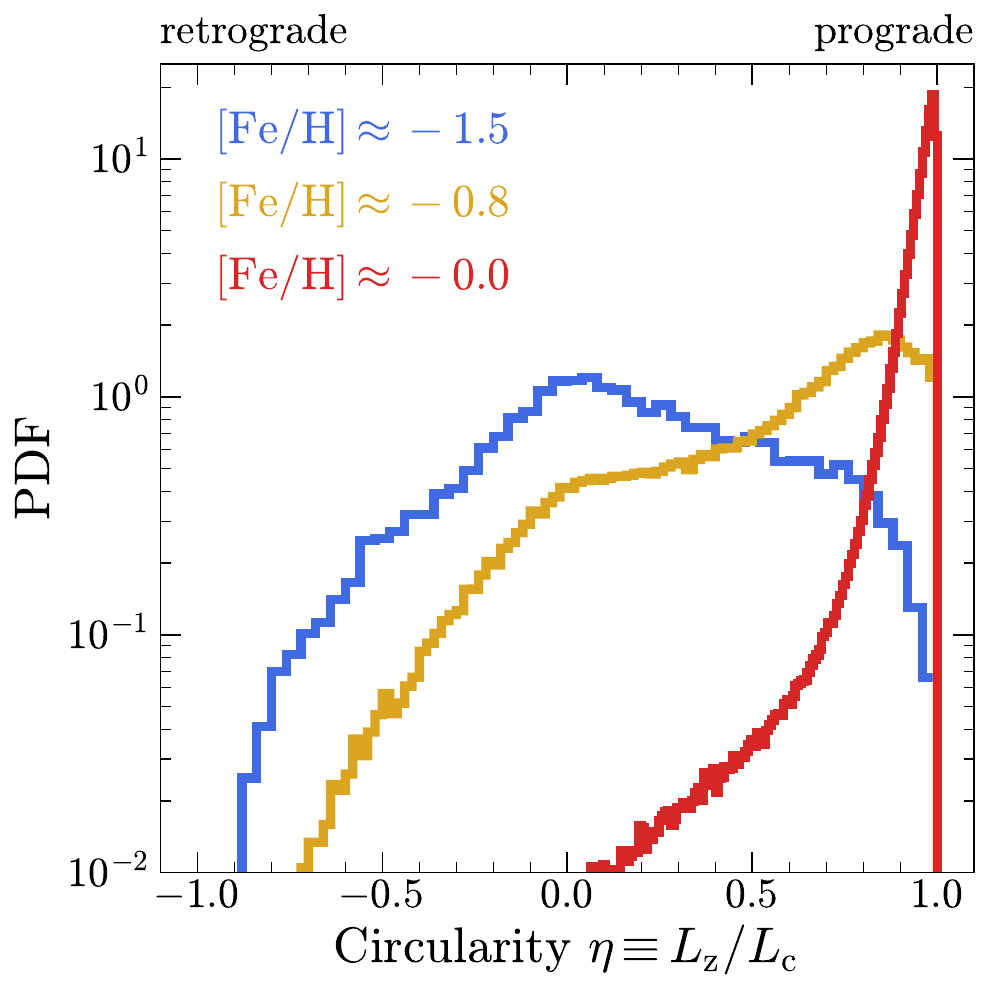}
    \caption{Distribution of orbital circularity $\eta$ for \textit{Gaia} DR3 giants in three different metallicity selections. Stars at solar metallicity \feh{}~$\approx 0$ are on overwhelmingly prograde and circular orbits with $\eta \sim 1$. At \feh{}~$\approx -0.8$ there is predominantly a kinematically hot disk component, along with an isotropic `in-situ' halo. At \feh{}~$\approx -1.5$ the isotropic component dominates, albeit with a prominent tail towards prograde orbits.}
    \label{fig:etapdf}
\end{figure}

Using the adopted potential, we compute the azimuthal angular momentum $L_\mathrm{c}$ and total energy $E$ of a perfectly circular orbit. We interpolate the resulting $L_\mathrm{c}(E)$ curve for each observed star's total energy and calculate the `orbital circularity' \LzLc{}. The circularity compactly represents the orbital properties of a star, ranging from -1 (perfectly retrograde, in-plane orbit) to 1 (perfectly prograde, in-plane orbit), with intermediate values corresponding to radial and/or polar orbits. The past literature has used various notations $\epsilon$ \citep[e.g.,][]{Abadi2003b,Yu2021,Yu2023} or $\eta$ \citep[e.g.,][]{Naidu2020} to describe this parameter, and $\eta \equiv$\,\LzLc{} is adopted in this work.
Figure~\ref{fig:etapdf} illustrates the overall circularity distribution of our sample. 

\section{The Birth of the Milky Way in Gaia DR3}\label{sec:birth}

\begin{figure*}
    \centering
    \includegraphics[height=7cm]{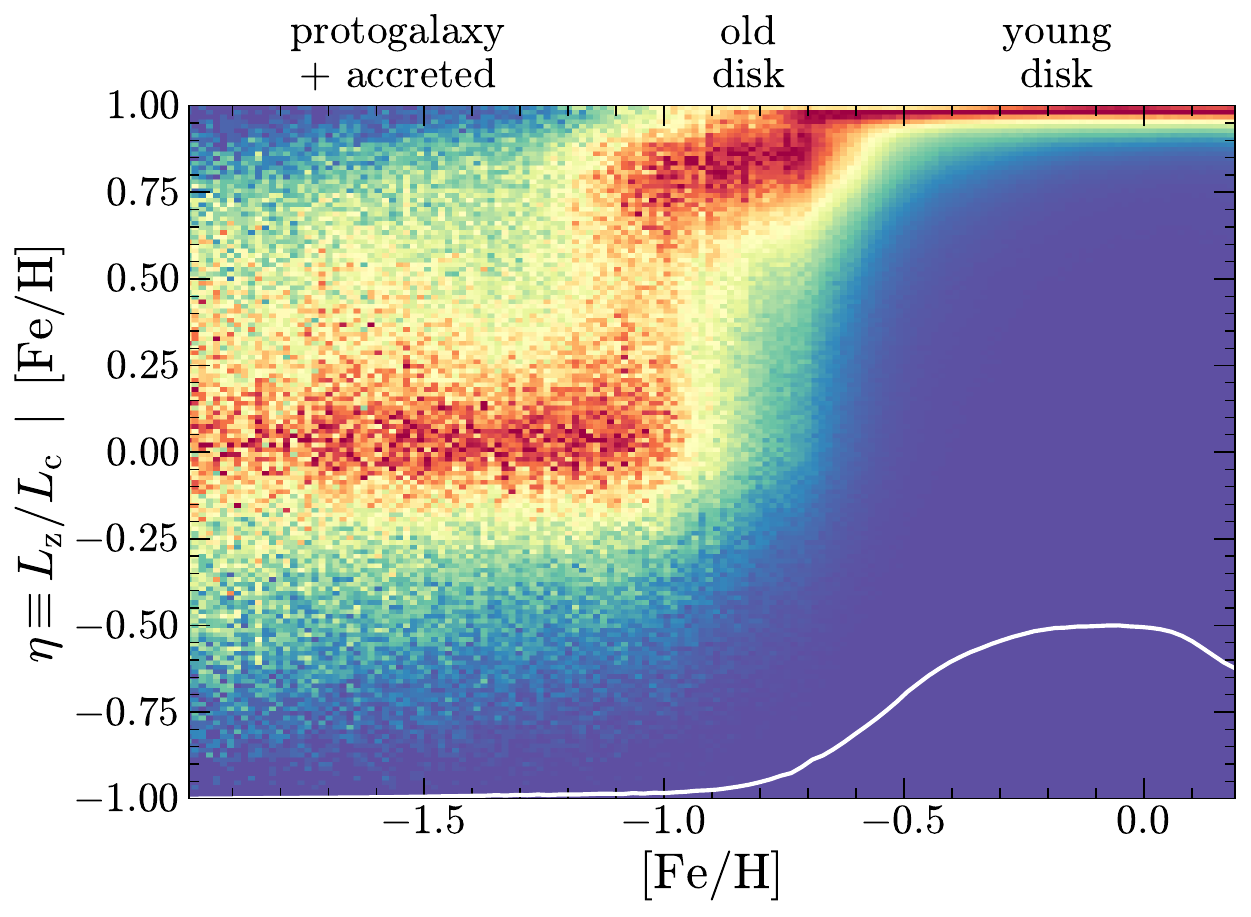}
    \includegraphics[height=6.5cm]{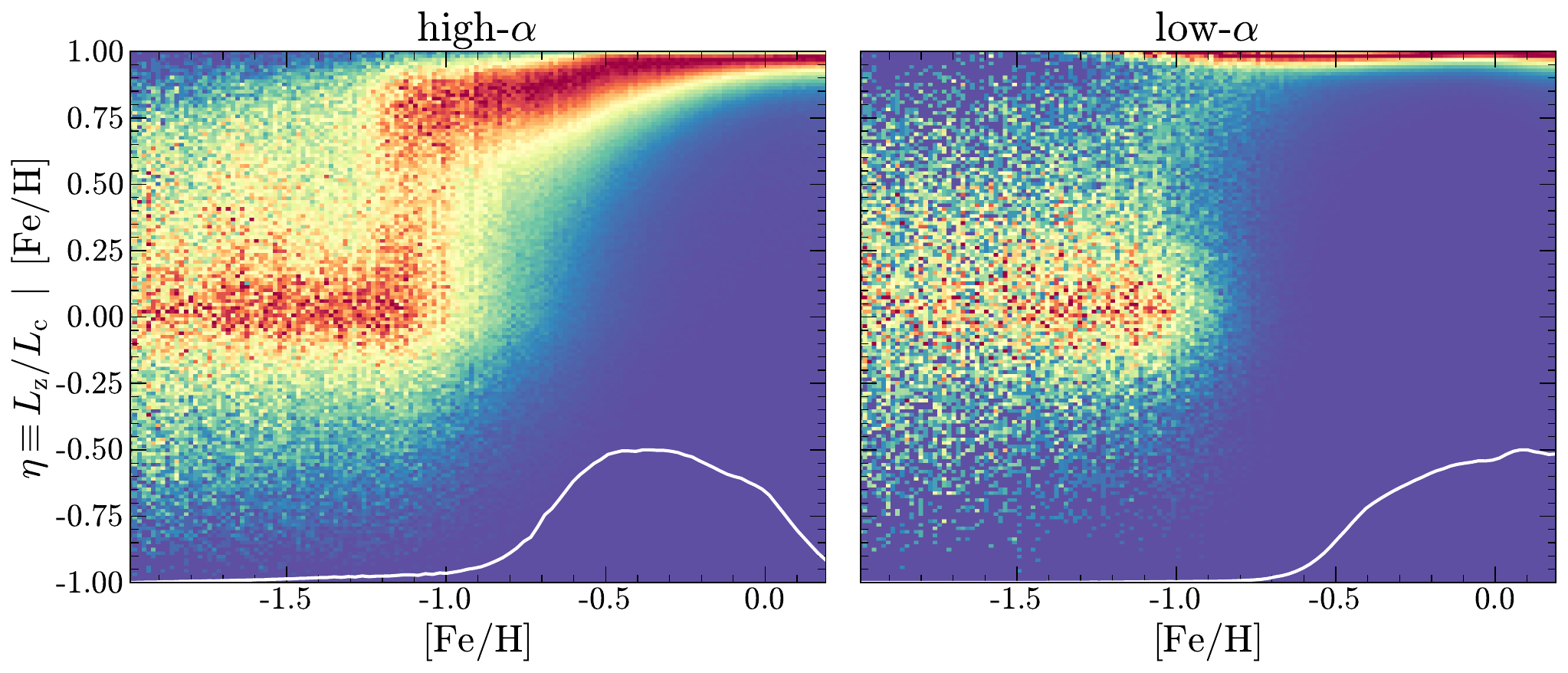}
    \caption{The three-phase evolution of the Milky Way revealed by our all-sky \textit{Gaia} giants. 
    The column-normalized distribution of orbital circularity is shown as a function of \textit{Gaia} XP metallicity --- a proxy for cosmic time. 
    The top panel shows all stars in our sample, whereas the bottom panels are split into high-$\alpha$ and low-$\alpha$ subsamples following Figure~\ref{fig:afe}. White lines show the metallicity distribution for each sample.
    [Fe/H] behaves as a particularly good cosmic clock for the high-$\alpha$ sequence in the bottom-left panel, since the age-metallicity relation is reasonably monotonic in this regime.}
    \label{fig:portrait}
\end{figure*}

Figure~\ref{fig:portrait} illustrates the formation and evolution of the Milky Way, showing the distribution of stars in angular momentum and metallicity, with metallicity serving as an imperfect proxy for cosmic time.
At each [Fe/H], Figure~\ref{fig:portrait} shows the column-normalized distribution of orbital circularity $\eta$. The top panel shows all stars in our RGB sample, whereas the bottom panels show the stellar populations differentiated by \afe{} following Figure~\ref{fig:afe}. 
Although it is tempting to read Figure~\ref{fig:portrait} as a temporal evolutionary sequence from left to right, the age-metallicity relation (AMR) of the Milky Way is neither linear, nor monotonic \citep[e.g.,][]{Strobel1991, Bensby2014, Feuillet2019, Sahlholdt2022, Xiang2022}. 
However, \citet{Xiang2022} have demonstrated that stars in the high-$\alpha$ disk obey a relatively tight and monotonic AMR. 
Therefore, the high-$\alpha$ panel of Figure~\ref{fig:portrait} can be interpreted as a temporal sequence, whereas the low-$\alpha$ panel is complicated by other effects such as the strong metallicity gradient. 
In the top panel of Figure~\ref{fig:portrait}, three distinct regimes are apparent.
In metallicity, the first transition is at  $-1.2 \lesssim \text{\feh{}} \lesssim -1.0$, while the second transition has substantive overlap around $-0.7 \lesssim \text{\feh{}} \lesssim -0.5$. 
Starting at low metallicity, we proceed with a discussion of these phases and transitions. 

\newpage
\subsection{The Protogalaxy}

Below [Fe/H]~$\approx -1.2$, the Galaxy is dominated by stars with minimal net rotation.
Interestingly, these stars are not entirely isotropic, and have a significantly non-zero median circularity of $\eta \approx 0.1$ in both the high-$\alpha$ and low-$\alpha$ samples. 
This population is likely dominated by protogalactic Milky Way stars, in addition to a significant fraction of stars accreted from past mergers like the Gaia-Sausage-Enceladus \citep[GSE;][]{Helmi2018,Belokurov2018a}. 
This interpretation is supported by the 
marginal distribution of \afe{} abundances at low metallicity, which for example at \feh{}~$\approx -1.6$ peaks sharply at \afe{}~$\approx 0.3$ (protogalactic), with a prominent tail towards \afe{}~$\approx 0.0$ (accreted).

Figure~\ref{fig:allsky} shows that these metal-poor stars cluster strongly in the Galactic center, as expected for the oldest stars formed in the MW \citep{Tumlinson2010, El-Badry2018, Rix2022, Horta2023, Semenov2023}.
From Figure~\ref{fig:wonky}, the circularity distribution of stars with \feh{}~$< -1.0$ is asymmetric, with a tail towards prograde orbits (i.e., $\eta > 0$). This feature was pointed out by \cite{Conroy2022}, who noted a net rotation in the H3 Survey's protogalactic population (also seen in the protogalactic population named `Aurora' by \citealt{Belokurov2022a}). 
Whether or not the protogalaxy had significant net rotation when its components first assembled remains uncertain. 
\cite{McCluskey2023} show that protogalactic populations with zero net rotational velocity $v_\phi$ at formation could acquire a net positive $v_\phi$ by the present day due to dynamical heating.
We explore this possibility in $\S$\ref{sec:tng:protogal} using the TNG50 simulations, affirming this mechanism in a MW analog. 


\subsection{Spinup and Cooldown: the Birth of the Disk(s)}
\label{sec:birth:disks}

At $-1.2 \lesssim \text{[Fe/H]} \lesssim -1.0$, there is a dramatic phase transition, with the emergence of a rotation-dominated disk component with $\eta \approx 0.75$, along with significant scatter. 
This is coincident with the `spinup' transition identified in $v_\phi$ by \cite{Belokurov2022a,Conroy2022}. 
It is apparent from the lower panels of Figure~\ref{fig:portrait} that the spinup transition is traced almost exclusively by high-$\alpha$ stars, whose AMR appears monotonic \citep{Xiang2022}: more metal rich means younger. 
The bottom-left panel of Figure~\ref{fig:portrait} therefore illustrates the birth of the Milky Way's disk over cosmic time, from the turbulent protogalaxy to an increasingly ordered disk. 
As metallicity (and time) increases, the range of circularities spanned by the high-$\alpha$ disk narrows, as stars move on increasingly circular orbits.
This `cooling down' of the orbital kinematics is notably smooth and continuous all the way up to solar metallicity, and reminiscent of the disk `settling' that is observed in external galaxies and simulations \citep[e.g.,][]{Kassin2012, Simons2017, Ma2017, Tiley2021, Gurvich2023}. 
Note that for the high-$\alpha$ disk, solar metallicity was reached about 8~Gyr ago \citep{Xiang2022}, and this panel therefore represents the first $\approx 5$~Gyr of the MW's evolution.

The top panel of Figure~\ref{fig:portrait} shows another distinct phase transition to yet more circular orbits, at \feh{}~$\approx -0.6$. 
This transition was noted by \cite{Belokurov2022a} as a narrowing of the $v_\phi$ velocity distribution as a function of metallicity. 
Most stars more metal-rich than this `cooldown transition' are on very cold and near-coplanar orbits, with $\eta \gtrsim 0.95$ and minimal scatter in $\eta$. 
The behavior of this transition becomes clearer when the sample is split by \afe{}: it is a transition from a high-$\alpha$ dominated disk to a low-$\alpha$ dominated disk. 
Indeed, the high-$\alpha$ disk continues to `cool down' smoothly from $-1.0 \lesssim$~\feh~$\lesssim 0.0$. 
In contrast, the low-$\alpha$ disk stars have nearly circular orbits across the full metallicity range \feh{}~$\gtrsim -0.6$. 
Here, it is important to emphasize that among low-$\alpha$ disk stars \feh{} is \emph{not} a proxy for temporal evolution. It mostly reflects the radial metallicity gradient: increasingly more metal-poor low-$\alpha$ disk stars reside at larger radii and angular momenta (Figure~\ref{fig:Rz}). 
In this context \feh{} may  reflect mostly the stars' birth radius rather than their age \citep[e.g.][]{Frankel2018,Frankel2020}. 

\begin{figure*}
    \centering
    \includegraphics[width=0.9\textwidth]{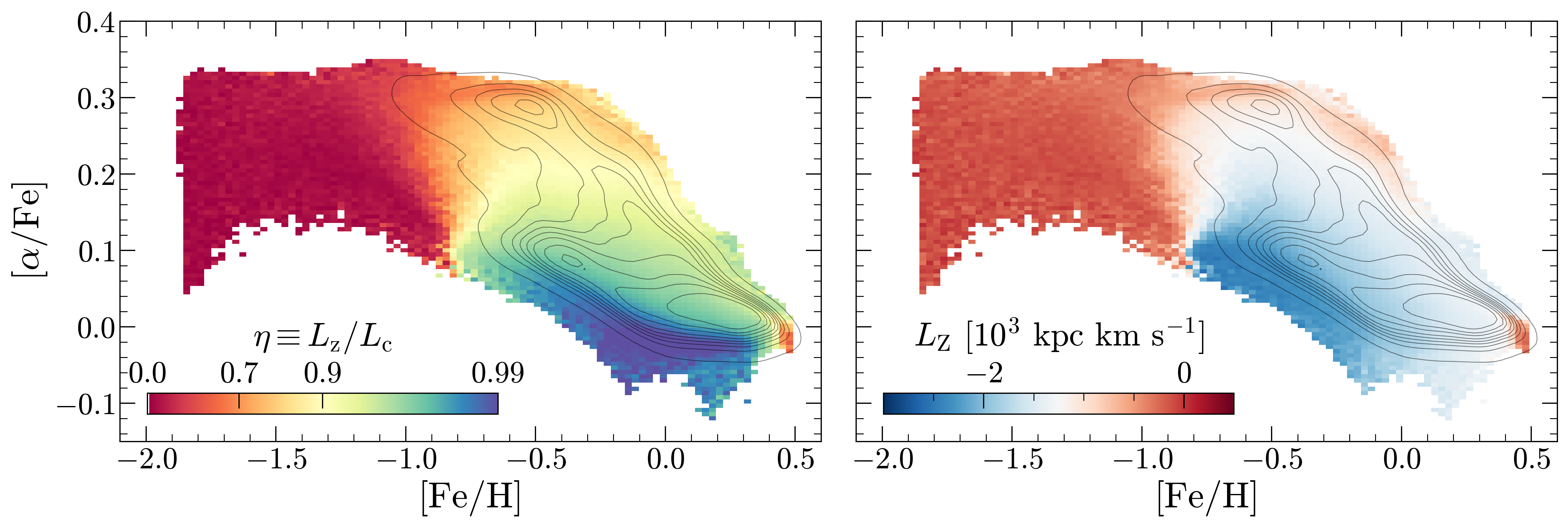}\vspace*{0cm}
    \includegraphics[height=1.1\columnwidth]{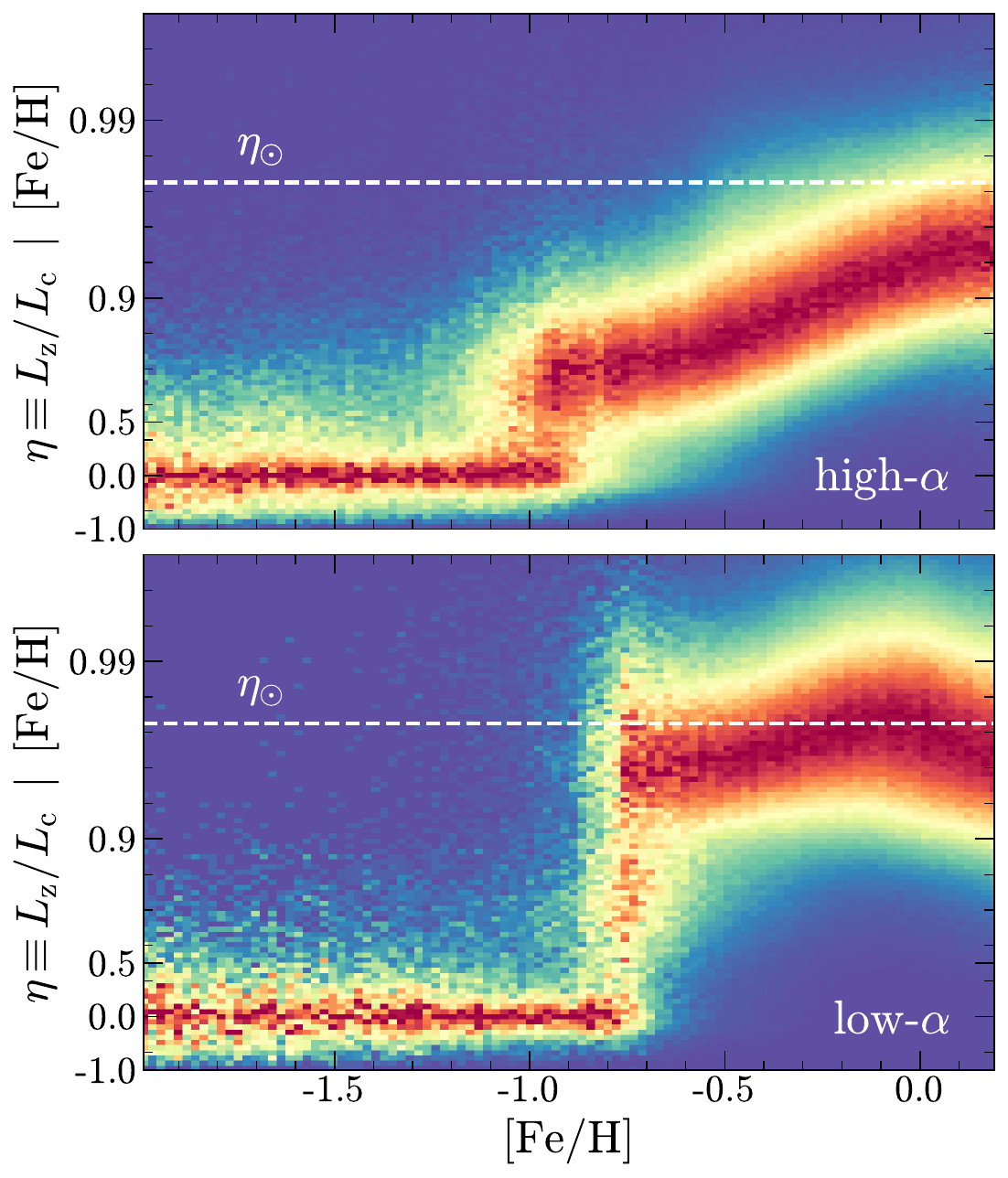}
    \hspace*{0.2cm}\includegraphics[height=1.1\columnwidth]{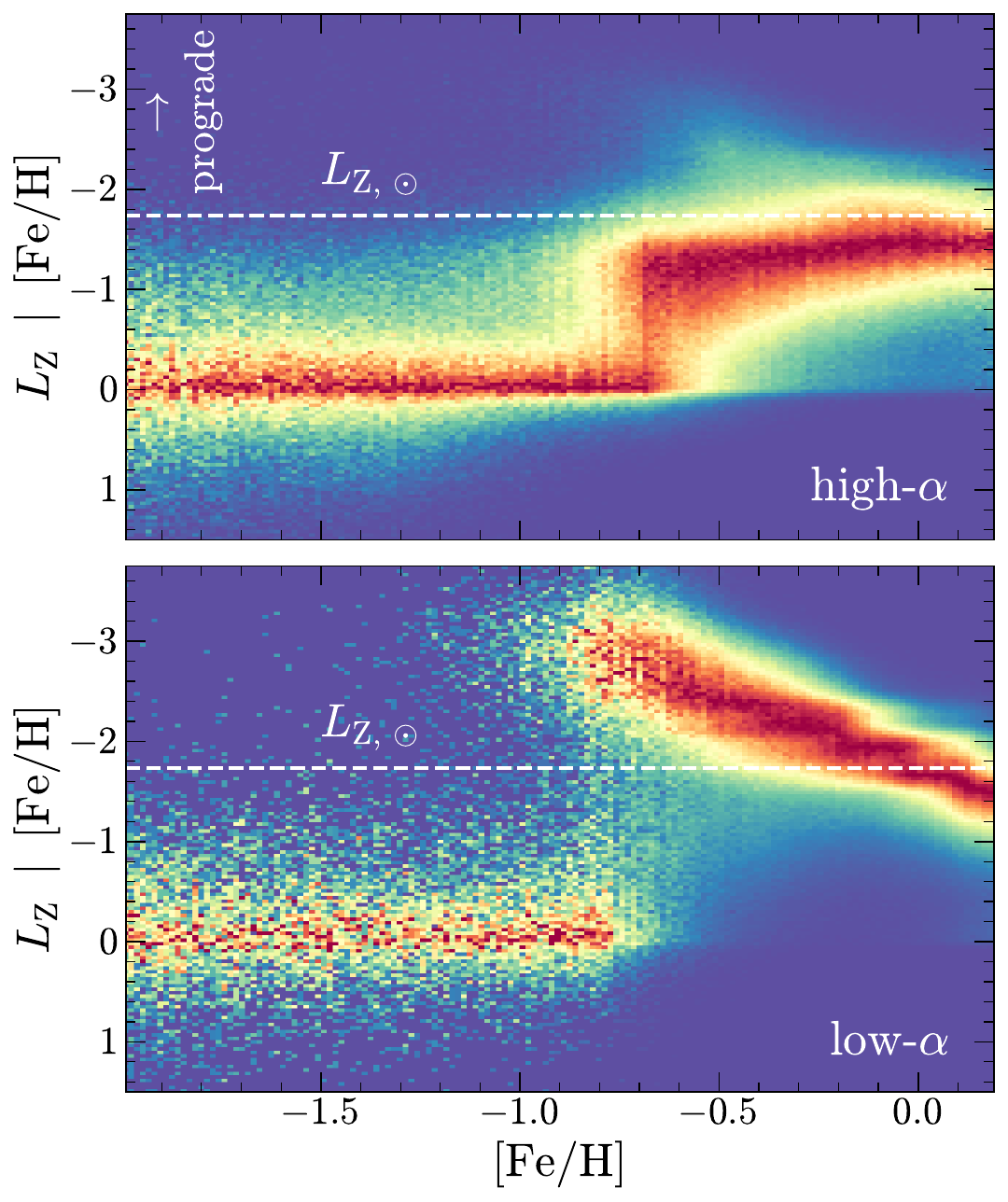}
    \caption{A summary of our kinematic results in the \feh{}-\afe{} plane. 
    The top panels show the median logarithmic circularity (left) and $L_\mathrm{Z}$ angular momentum  (right) respectively, and the lower panels show column-normalized profiles of these quantities versus metallicity, split by \afe{} following Figure~\ref{fig:afe}.
    To guide the eye, a dashed white line shows the characteristic circularity and angular momentum for the low-$\alpha$ disk at solar metallicity.
    }
    \label{fig:wonky}
\end{figure*}

To zoom into these later phases in the Galactic disk(s) evolution, it is useful to apply a non-linear transformation to the circularity $\eta$, using $\log{(1-\eta)}$ instead, as shown in Figure~\ref{fig:wonky}. 
This `logarithmic circularity' maps monotonically to $\eta$ but visually expands the dynamic range among stars on nearly circular and co-planar orbits, making it a more suitable metric to chart out the $\eta \gtrsim 0.9$ disk regime. 
Note that Figure~\ref{fig:wonky} is quite sensitive to the dust-modulated selection effect that limits our sample's depth at low Galactic latitudes (see Figures~\ref{fig:allsky},\ref{fig:Rz}). 
That is, stars with $\eta \gtrsim 0.99$ likely exist throughout the cold disk of the MW at very low $|b|$, but are hidden behind dust---especially at low Galactocentric radii---and are consequently missing from Figure~\ref{fig:wonky}.

Figure~\ref{fig:wonky} illustrates the  markedly different distributions of the high-$\alpha$ and low-$\alpha$ disks between $-0.7 \lesssim$~\feh{}~$\lesssim 0.0$, with the low-$\alpha$ disk residing on kinematically colder orbits throughout this range. 
As the trends in Figure~\ref{fig:wonky} illustrate, the low-$\alpha$ stars show a sharp break in circularity at \feh{}~$\sim -0.8$. This seems to imply that there is no low-$\alpha$ disk below \feh{}~$\sim -0.8$; the isotropic ($\eta \sim 0$) stars in this panel reflect the accreted halo population, which is metal-poor and has low $\alpha$. 
As noted above, the high-$\alpha$ disk has a steady and log-linear circularity evolution all the way from $-1.0 \lesssim$~\feh~$\lesssim 0.0$. 

The kinematic distinctions between the high-$\alpha$ and low-$\alpha$ disks are emphasized in the top panel of Figure~\ref{fig:wonky}, which displays the $L_\mathrm{Z}$ angular momentum (left) and logarithmic circularity (right) in the \afe{}-\feh{} space for our entire sample. 
The top-right panel of Figure~\ref{fig:wonky} illustrates how the concepts of `hot disk' and `cold disk'  map onto the `high-$\alpha$' and `low-$\alpha$' disks, respectively. 
As seen in Figure~\ref{fig:wonky}, the high-$\alpha$ disk exhibits a gradual cooling down of orbits as metallicity increases, whereas the apparent trend in the low-$\alpha$ disk is primarily set by the radial gradient and selection function. 
The most circular stars in the outer disk are obscured by dust in the mid-plane, decreasing the median circularity of the outer disk. 
It will be a worthwhile endeavour for future work to disentangle these effects by modelling the selection function directly. 

\subsection{The Heated Old Disk and In-Situ Halo}

\begin{figure}
    \centering
    \includegraphics[width=\columnwidth]{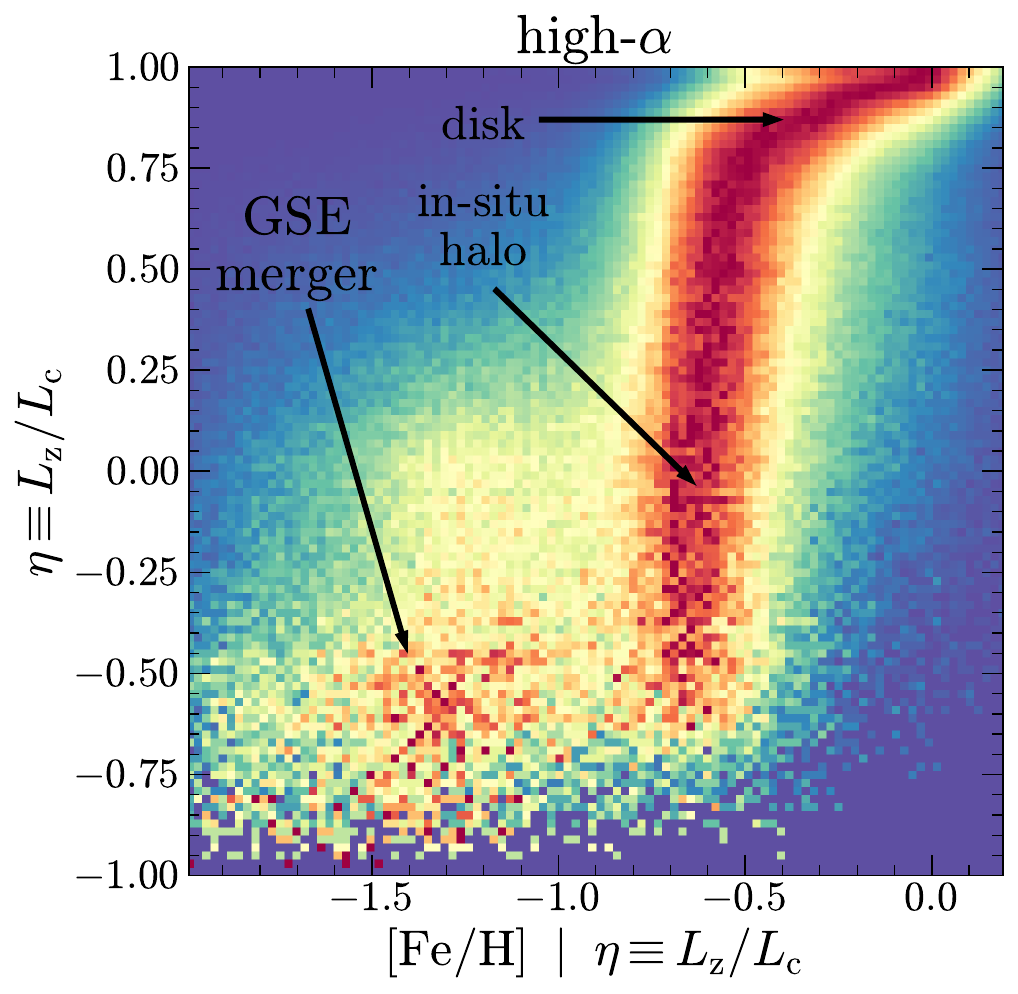}
    \caption{Row-normalized version of the high-alpha panel from Figure~\ref{fig:portrait}---the conditional metallicity distribution for high-alpha stars at each orbital circularity.
    Three key structures are highlighted: the disk itself, the in-situ halo, and accreted stars from the GSE merger that persist even in the high-$\alpha$ sample. 
    }
    \label{fig:splash}
\end{figure}

The protogalactic and hot disk components overlap between $-1.2 \lesssim \mathrm{Fe/H} \lesssim -0.9$ in Figure~\ref{fig:portrait}. In this region, there is also a swath of stars at intermediate orbits between the two populations, that visibly extends up to [Fe/H]~$\approx -0.5$. 
This entire sequence coincides with the known `in-situ halo' or `splash' population \citep{Bonaca2017, Belokurov2020b}. 
To visualize this portion of the distribution more clearly, we display the row-normalized metallicity-circularity distribution in Figure~\ref{fig:splash}. 
This plot shows the metallicity distribution function at each value of orbital circularity. 
The most circular high-alpha stars ($\eta > 0.75$ and [Fe/H] $>-0.5$) exhibit a continuous evolution towards lower metallicities, with a break at \feh{}~$\approx -0.5$. 
Below this metallicity, high-alpha stars occupy a broad swath of orbits.
These stars have metallicities identical to the thick (or kinematically hot) disk, but extend all the way to isotropic and even retrograde orbits. 

The most plausible origin for these stars is that they were born in the early MW disk and were subsequently kicked up onto isotropic and eccentric orbits by a major merger, presumably GSE itself \citep{Bonaca2017, Belokurov2020b}. 
Indeed remnant stars from the GSE merger are clearly visible in Figure~\ref{fig:splash} as an independent swath of isotropic-to-retrograde stars at \feh{}~$\approx -1.2$. 

\cite{Belokurov2020b} find that the age distribution of isotropic `splashed' stars truncates sharply $\approx 9.5$~Gyr ago, potentially timing the merger's impulse on the old disk. 
This is consistent with the transition metallicity of \feh{}~$\approx -0.5$ seen in Figure~\ref{fig:splash}: stars formed before this metallicity are heated into the in-situ halo, whereas stars formed after this metallicity continue the settling of the old disk towards more circular orbits. 
In $\S$\ref{sec:tng:olddisk} we use TNG50 simulations to lend credence to this hypothesis, finding that the kinematic `hotness' of both the old disk and the in-situ halo population can be created by a single major merger.

\section{Birth of a MW analog in TNG50}\label{sec:tng}

In $\S$\ref{sec:birth}, we used \textit{Gaia} DR3 to unveil a comprehensive perspective on the structure and kinematics of the Milky Way. 
We are now in a better position to hunt in large-scale cosmological simulations for \textit{evolutionary} analogs of the MW, rather than simply structural ones that match its present-day properties.
As stated earlier, the eventual goal is to produce a plausible biographical story of our own Galaxy's birth and evolution. 

In this section, we search the TNG50 cosmological simulation for analogs of the Milky Way that match the \textit{Gaia} data presented here.
We build on the selections of \citet[][hereafter \citetalias{Semenov2023}]{Semenov2023}, who isolated MW-like galaxies in TNG50 that exhibit the `early spin up' in $v_\phi$ as a function of metallicity \citep{Belokurov2022a}. 
TNG50 is the highest-resolution box from the IllustrisTNG suite of hydrodynamical cosmological simulations \citep{Pillepich2018, Nelson2019, Weinberger2020}. 
The simulation spans $50^3$ co-moving Mpc$^3$ with a dark matter mass resolution $4.5 \times 10^5\Msun$, and a baryon mass resolution of $\sim 8.5 \times 10^4\Msun$. 
For more details on the initial conditions, sub-grid physics, and implementation, we defer to the above IllustrisTNG references. 

\begin{figure}
    \centering
    \includegraphics[width=\columnwidth]{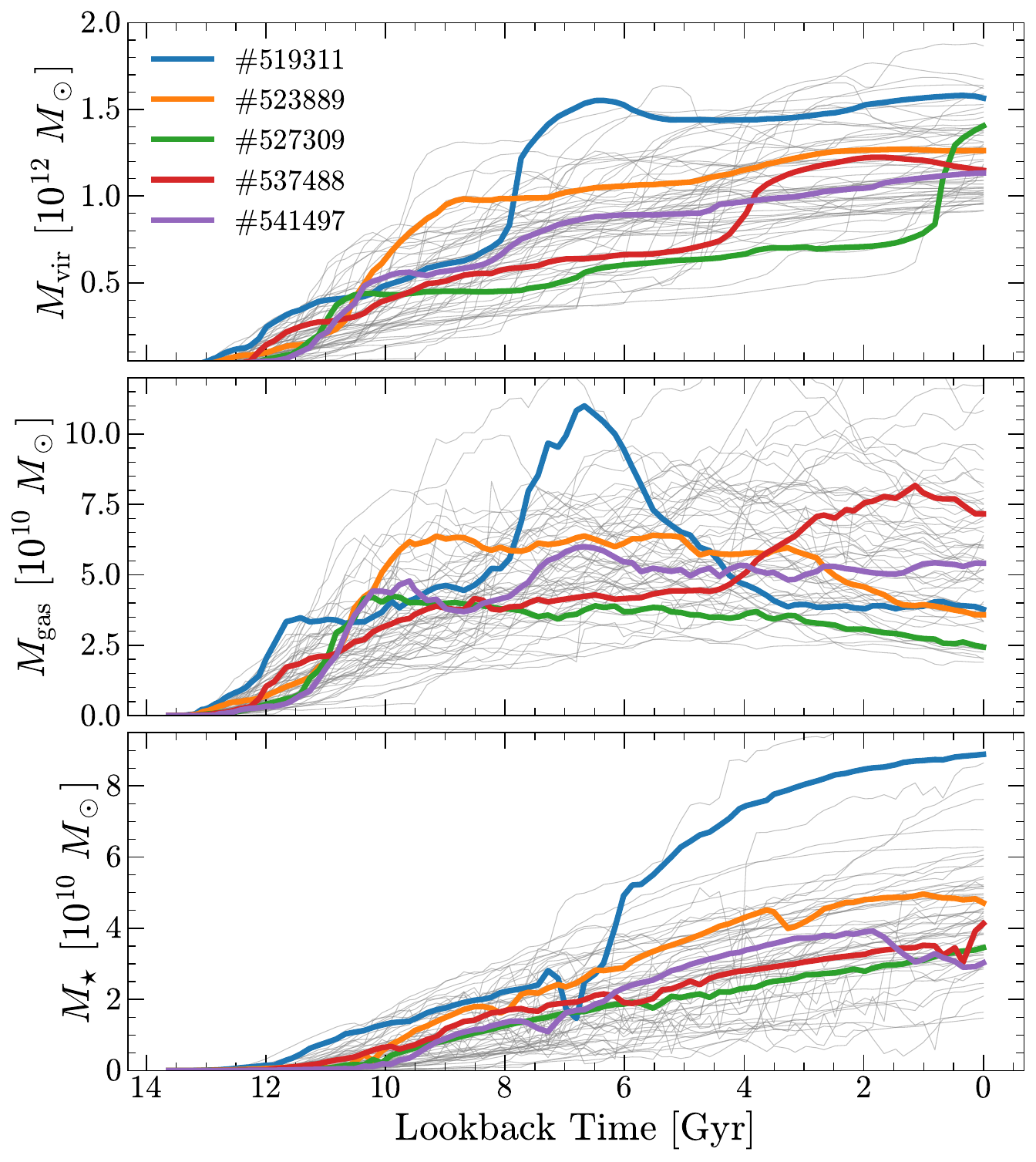}
    \caption{Mass accretion history for 61 MW-like TNG50 halos (gray lines) selected by \citetalias{Semenov2023}, showing the virial mass (top), gas mass (middle), and stellar mass (bottom). 
    The 5 early-spinup halos analyzed in this work are highlighted with colored lines.}
    \label{fig:tng_mah}
\end{figure}

\begin{figure*}
    \centering
    \includegraphics[width=\textwidth]{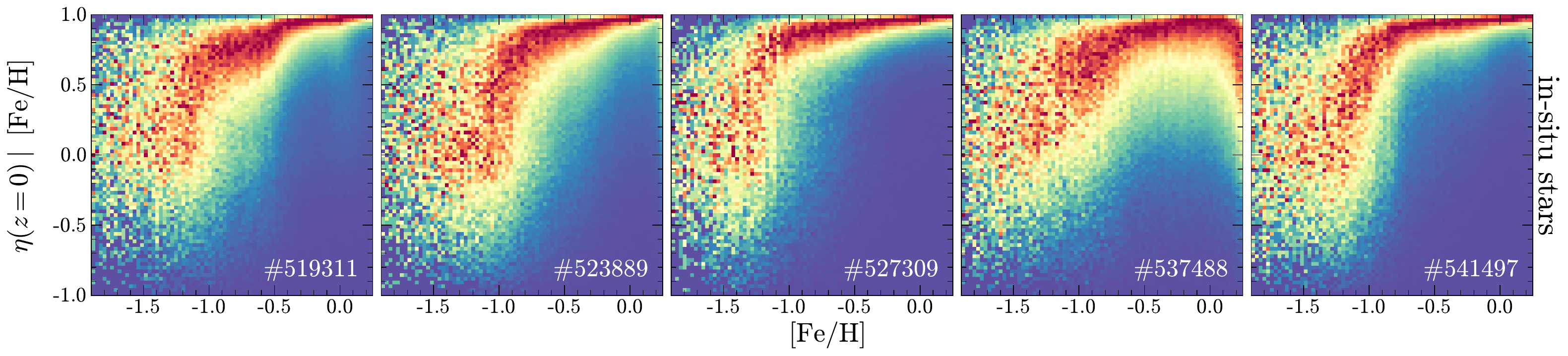}\vspace{-0.1cm}
    \includegraphics[width=\textwidth]{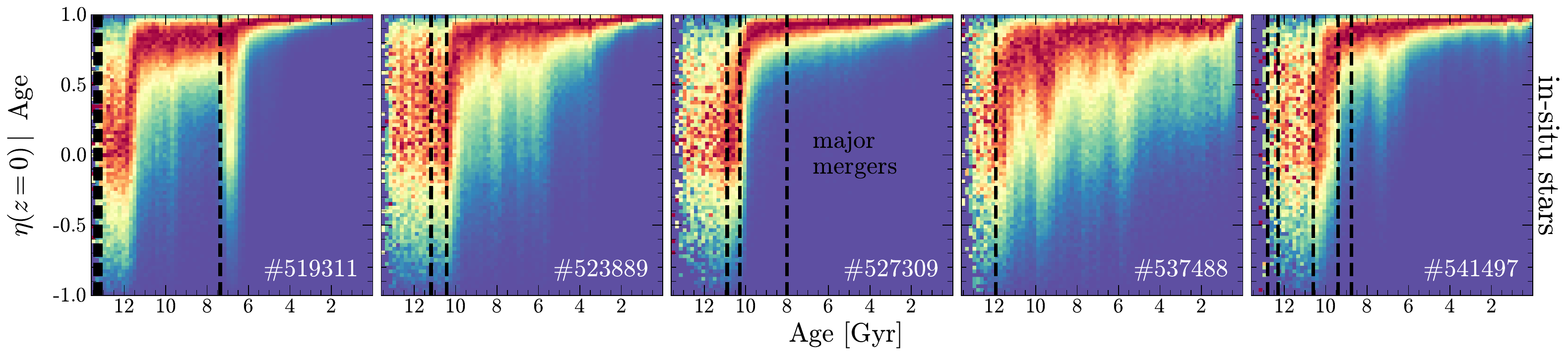}
    \caption{The evolution of orbital circularity in five `early spinup' MW analogs from the TNG50 simulations. Only \textit{in-situ} stars born within 30~kpc (comoving) of the primary halo are shown.
    Top: Column-normalized distribution of orbital circularity as a function of metallicity. 
    Note the similarity between TNG50 galaxy \#519311 and the MW data shown in Figure~\ref{fig:portrait}. 
    Bottom: Column-normalized distribution of orbital circularity as a function of stellar age. 
    Vertical lines denote snapshots containing relatively major mergers, where the accreted stellar mass increases by $> 5\%$.}
    \label{fig:tng_portrait}
\end{figure*}

Our full sample of simulated MW analogs \citepalias{Semenov2023} consists of 61 star-forming galaxies with halo mass (within the spherical overdensity of 200$\times$critical density of the universe) of $M_{\rm 200c} = 0.8\text{--}1.4 \times 10^{12}\Msun$ and $\SFR > 0.2\Msunyr$ that exhibit rotationally supported stellar disks with $\vrot/\sigma_v > 6$ for stars younger than $100\Myr$. 
By examining the evolution of the azimuthal Galactocentric velocity $v_\phi$ as a function of metallicity, \citetalias{Semenov2023} identified 11 TNG50 galaxies that matched the `early spin-up' reported in MW data by \cite{Belokurov2022a}. 
These galaxies all have well-settled, extended disks at $z = 0$, in contrast with the `late spin-up' galaxies which are less disk dominated at the present day, with less ordered orbits.
\citetalias{Semenov2023} report that the key evolutionary feature distinguishing these populations is that the early spin-up galaxies undergo more rapid mass assembly at early times. 
We note that although \citetalias{Semenov2023} used a CDF-matching approach to rescale their simulation metallicities to the \cite{Belokurov2022a} data, our \textit{Gaia} data are subject to a very different selection function.
We therefore directly use the metallicities reported by the simulation, which are sufficient for the qualitative analysis presented here. 
Figure~\ref{fig:tng_mah} shows the dark matter, gas, and stellar mass accretion history of five of the galaxies that spun up earliest, which are the ones we analyze further in this work. 

\subsection{The Evolution of Spin in TNG50 Disks}

The kinematic evolution of our five early-spinup TNG50 galaxies is shown in Figure~\ref{fig:tng_portrait}. 
Only `in-situ' stars, those born within $30$~kpc (comoving) of the galaxy, are shown. 
The top row corresponds to our Figure~\ref{fig:portrait}, showing the column-normalized distribution of circularity $\eta$ versus metallicity. 
Remarkably, the galaxy in halo \#519311 exhibits the same distinctive three-phase behaviour seen in our MW data (Figure~\ref{fig:portrait}).
We have verified that \#519311 is the most similar to the Milky Way's  $\eta$--\feh{} data among all 61 MW-mass halos identified by \citetalias{Semenov2023}. 

The bottom row in Figure~\ref{fig:tng_portrait} provides the much-needed dimension that our data in Figure~\ref{fig:portrait} are lacking: precise stellar ages. 
These circularity diagrams, $\eta (\mathrm{Age})$ can genuinely be interpreted as evolutionary sequences that show what orbits stars born at different epochs end up on. 
We emphasize that $\eta$ here denotes the orbital properties at $z = 0$, not at birth. 
As a consequence of the selection in \citetalias{Semenov2023}, all these galaxies `spin up' more than 10~Gyr ago, matching MW data.
Several sharp spikes towards isotropic orbits are visible in most halos, which coincide with merger events. 
Dashed lines indicate `major merger' events, which are selected to increase the fraction of accreted stellar mass in the galaxy by $> 5\%$. 
Although this metric is imperfect---as evidenced by smaller mergers or gas accretion events without preceding dashed lines---it is sufficient to highlight the most prominent mergers in a galaxy's history.  

\begin{figure*}
    \centering
    \hspace*{-0.2cm}\includegraphics[width=\textwidth]{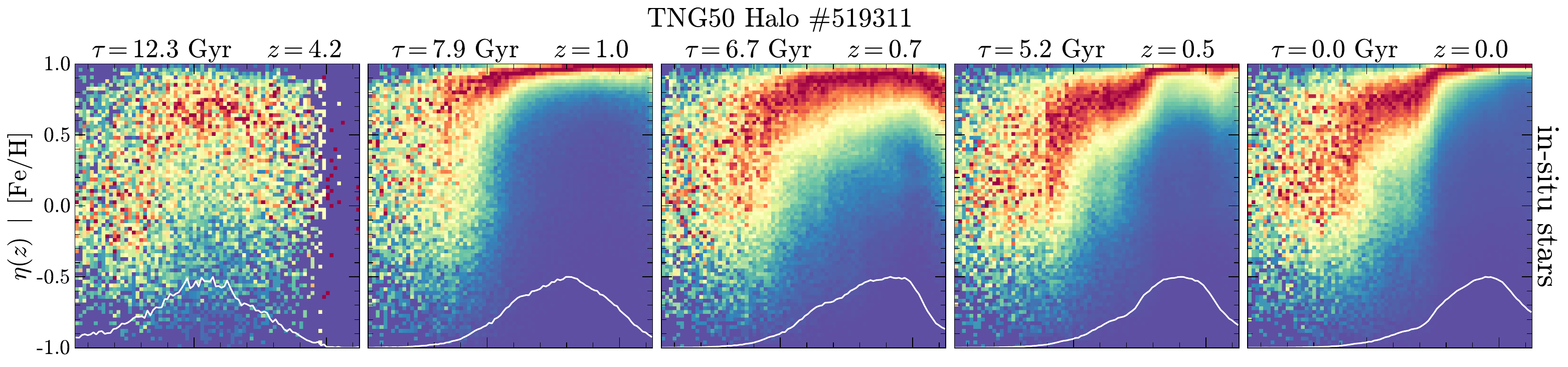}\vspace{-0.1cm}
    \hspace*{-0.2cm}\includegraphics[width=\textwidth]{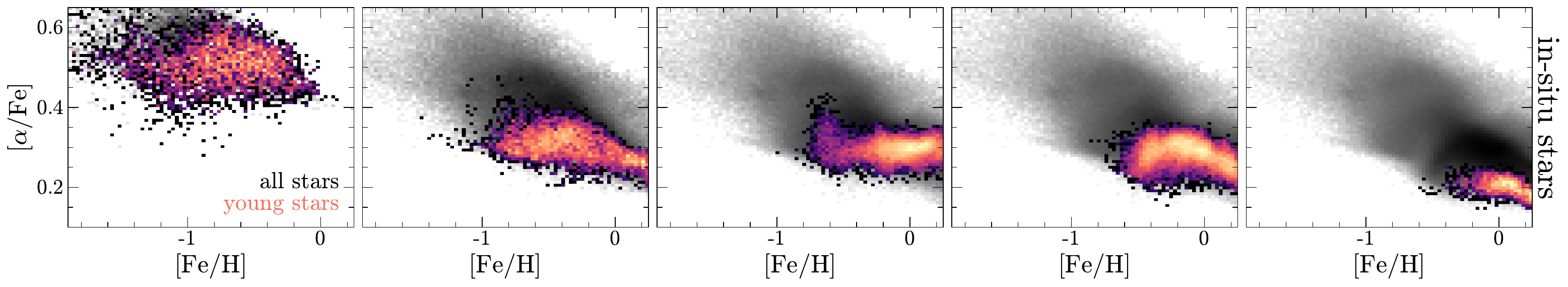}
    \hspace*{0.15cm}\includegraphics[width=0.95\textwidth]{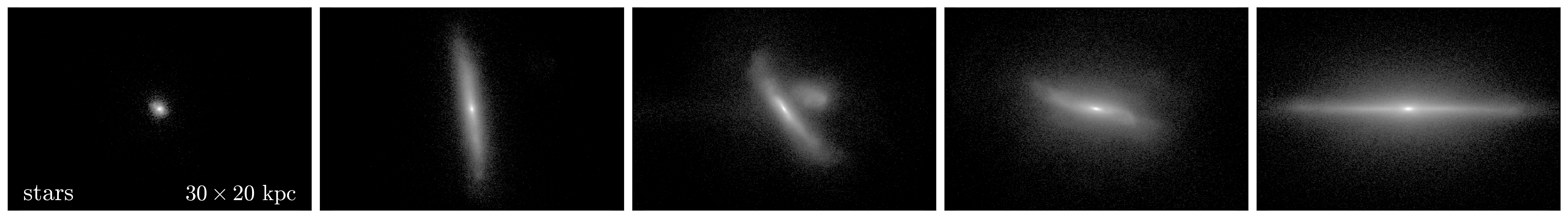}
    \hspace*{0.15cm}\includegraphics[width=0.95\textwidth]{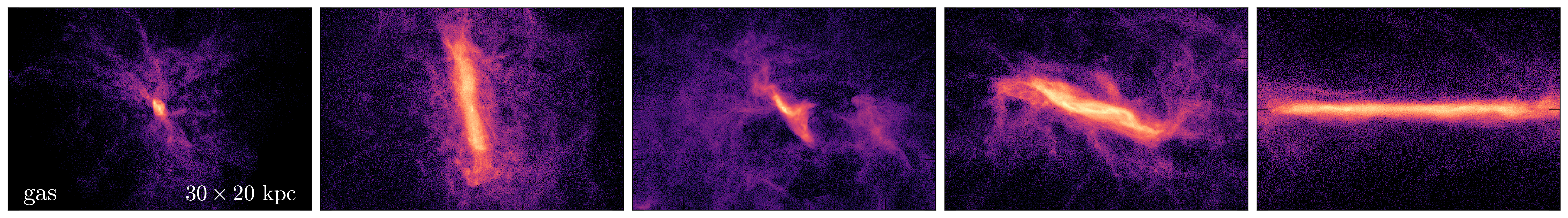}
    \caption{The evolution of TNG50 halo \#519311 over cosmic time. 
    5 key snapshots are temporally ordered from left to right, with the lookback time and redshift indicated at the top. 
    The top panel shows the orbital circularity-metallicity space for all \textit{in-situ} stars born up to that time, whereas the second panel shows the \feh{}-\afe{} plane, with stars younger than 100 Myr in each snapshot highlighted.
    The orbital circularities are computed in a coordinate frame aligned with the net angular momentum of each snapshot.
    The third panel shows the logarithmic density of star particles in fixed coordinates oriented edge-on to the present-day disk, and the fourth panel shows the same for gas particles.
    This halo undergoes a major merger at $z \approx 0.7$, visible in the central panels.
    An animated version of this figure is available \href{https://youtu.be/1q1TLoLUqdY}{here}.
    }
    \label{fig:eta_evol}
\end{figure*}

Interestingly, in the age--$\eta$ space, the three-stage picture of disk evolution (protogalaxy, then old hot disk, then young cold disk) is evident for all halos except \#527309, while in the [Fe/H]--$\eta$ space, the distinction between old hot and young cold disk is less apparent. 
This is a consequence of a significant overlap in the chemical evolution of the two disks (see Section~\ref{sec:birth:disks}), implying that, while the three-phase formation picture can be common, its imprint on the chemo-kinematic distribution of stars is less common.
The heated old disk population is typically superseded by the cold, young disk across all metallicities. 
However, in halo \#519311 there is enough of a metallicity difference between the disks for the old disk to remain visible in the [Fe/H]--$\eta$ space at the present day, matching the three-phase picture observed in the MW.

\subsection{Halo \#519311: An Illustrative Milky Way Analog}
\label{sec:tng:analog}

As mentioned above, TNG50 halo \#519311 exhibits a `three-phase' structure in the circularity-metallicity space that is remarkably similar to the MW data (compare Figure~\ref{fig:tng_portrait} to Figure~\ref{fig:portrait}). 
This halo is one of the earliest-spinup galaxies in the \citetalias{Semenov2023} sample, and also experiences some of the earliest mass growth in the entire TNG50 MW analog sample (Figure~\ref{fig:tng_mah}). 
This particular halo also experiences a uniquely gas-rich major merger $\approx 8$~Gyr ago (bottom panel of Figure~\ref{fig:tng_mah}). 
Although this merger is slightly later than the hypothesized timing of the Milky Way's own GSE merger \citep{Bonaca2020, Belokurov2020b, Naidu2021}, the fact that \#519311 is the only MW-like TNG50 galaxy to show the distinctive three-phase kinematic behavior in the circularity-metallicity space warrants further investigation. 
In this section, we take a closer look at halo \#519311 as a Milky Way analog. 

Figure~\ref{fig:eta_evol} illustrates the evolutionary history of TNG50 halo \#519311. 
Although the $\eta (\mathrm{Age})$ shown in the bottom panels of Figure~\ref{fig:tng_portrait} helps to sort the present-day orbits of stars, it lacks information about the \textit{birth} orbits. 
In the top row, each panel shows what the $\eta (z | \mathrm{\feh{}}$) distribution would have looked like at different epochs. 
The orbital circularities $\eta(z)$ are computed at that epoch for all \textit{in-situ} stars born to that point. 
The second row illustrates the \feh{}-\afe{} abundance plane for all stars born up to that point (grey) and for young stars born within 100~Myr of the chosen snapshot.
The lower two rows show the distribution of stars and gas in each snapshot, in a fixed coordinate frame chosen to be the `edge-on' orientation of the disk at $z = 0$. 
Note that the circularity $\eta(z)$ is not computed in this fixed frame, but in a frame aligned with the net angular momentum vector of the galaxy in each snapshot.
Between the second and third panels from the left, this halo experiences a major merger at $z \approx 0.7$ ($\tau \approx 7$~Gyr). 
The accreted dwarf galaxy is visible in the stellar and gas images in the central panel. 

At $z = 4.2$, a billion years after the Big Bang, a nascent rotating population of relatively high \feh{} has started to succeed the isotropic protogalaxy. 
Yet, as the images in Figure~\ref{fig:eta_evol} show, the galaxy is still spheroidal and clumpy, fed by numerous chaotic streams of gas. 
By $z = 2$, the spinup transition is complete and a remarkably thin, kinematically coherent disk has already formed. 
At the time of its formation, this old disk was kinematically much colder than it is now, with $\eta \gtrsim 0.9$ across a wide range of metallicities. 
This suggests a significant amount of kinematic heating, the source of which is immediately apparent in the next snapshot at $z \approx 0.7$: the arrival of a major gas-rich merger. 

The major merger is visible in the central panels of Figure~\ref{fig:eta_evol}, both in the distribution of accreted stars and in the images of the stellar and gas density. 
Several interesting features can be gleaned from these panels alone. 
We emphasize that this is an exceptionally major merger among the early-spinup TNG50 galaxies we analyze. 
As Figure~\ref{fig:tng_mah} illustrates, among all our MW analogs, halo \#519311 
experiences one of the most massive and gas-rich mergers that does not destroy the preexisting stellar disk.
At the time of the merger, the distribution of circularities in the old disk thickens considerably. 
This is not only due to orbital heating by the merger but also due to a rotation of the disk itself: as can be seen by comparing the stellar images, the galaxy has tipped by $\approx 30^\circ$ immediately after the merger.
Indeed, we find that the merger and subsequent gas accretion contribute significant angular momentum to the galaxy, setting the final orientation of the galactic disk. 

A comparison between the stellar and gas distributions in the central panel of Figure~\ref{fig:eta_evol} also reveals that the accreted gas lags behind the accreted stars by several kpc. 
Although stars from the accreted dwarf have already reached the disk of the primary galaxy, the bulk of the accreted gas is still $\approx 10$~kpc above the disk. 
Although a precise investigation of this offset is warranted, a likely explanation is tidal stripping, along with ram pressure stripping of the accreted dwarf by the circumgalactic medium of the primary galaxy \citep[e.g.,][]{Simpson2018, Rohr2023}. 
This additional drag force on the gas would naturally cause it to lag behind the collisionless stellar component. 

Moving forward to one Gyr after the merger, the galaxy remains in a state of significant disequilibrium. 
Although the stellar component of the accreted dwarf galaxy has started isotropizing into a spheroidal `in-situ halo,' there are still turbulent flows of gas accreting onto the disk. 
The rotation and torquing of the galaxy continues, now over $\approx 60^\circ$ from the initial orientation of the old disk. 
This inflow of gas also increases the gas density in the disk, triggering an intense period of star formation. 
This creates stars on exceptionally circular orbits starting at \feh~$\approx -0.5$. 
The $z = 1.0$ snapshot demonstrates that the old disk rapidly enriches to solar metallicities. 
Subsequently (see the $z = 0.5$ snapshot), it becomes superseded by the young disk at \feh{}~$\gtrsim -0.5$, creating the characteristic `three-phase' appearance of the metallicity-circularity plot. 

The last 5~Gyr of this galaxy's evolution are relatively quiescent, with the rest of the kinematically cold young disk assembling and enriching up to super-solar metallicities. 
The final orientation of the young disk is misaligned from the original old disk by almost $90^\circ$. 
The final metallicity-circularity distribution of this galaxy is remarkably similar to the MW data shown in Figure~\ref{fig:portrait}, a unique feature among all the early-spinup TNG50 galaxies (Figure~\ref{fig:tng_portrait}). 
This feature is rather unusual in the [Fe/H]--$\eta$ plane likely because of the relatively high metallicity of the young disk (\feh{}~$\gtrsim -0.5$), such that the remnants of the old disk with \feh{}~$\lesssim -0.5$ remain exposed in this plane. 
In the other examples in Figure~\ref{fig:tng_portrait}, the old disk is buried (in number counts) under the low-[Fe/H] end of the young disk.


\subsection{The Protogalaxy: Past and Present}\label{sec:tng:protogal}

\begin{figure}
    \centering
    \includegraphics[width=\columnwidth]{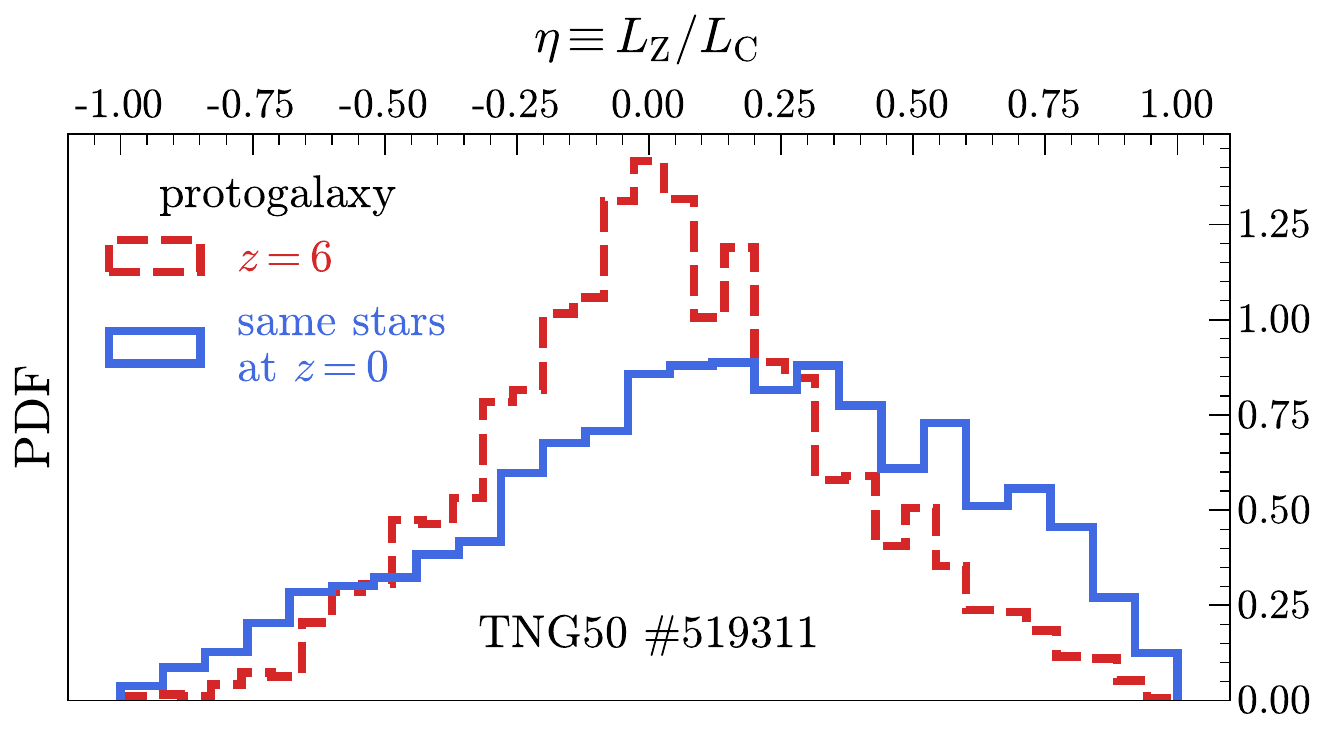}
    \includegraphics[width=0.93\columnwidth]{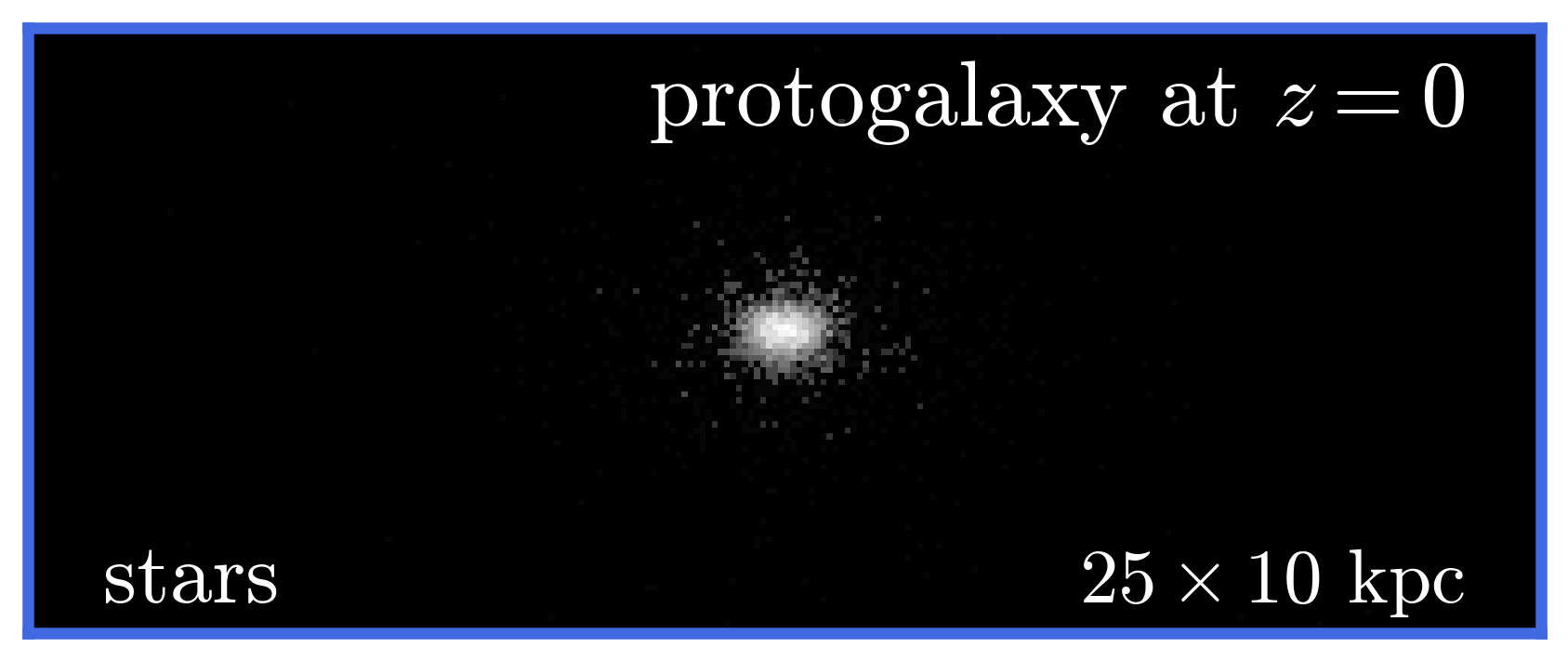}
    \caption{Top: The circularity distribution of all protogalactic stars in TNG50 halo \#519311 born up to $z = 6$ (red), and the circularity distribution of those same stars at the present day (blue).
    Bottom: Edge-on spatial distribution of protogalactic stars at the present day. 
    }
    \label{fig:519311_protogalaxy}
\end{figure}

Before the galaxy in TNG50 halo \#519311 begins the spinup transition to form the old disk, it assembles a sizable spheroidal `protogalaxy' that already contains $3 \times 10^8\,M_\odot$ of stellar mass by $z \approx 6$. 
\cite{Rix2022} utilized all-sky \textit{Gaia} data to unveil a substantial population of metal-poor stars residing in the heart of the MW, and used rough scaling arguments to estimate that the MW assembled $\gtrsim 5 \times 10^7\,M_\odot$ in stars by $z \gtrsim 5$. 
\cite{Belokurov2023} also estimate a stellar mass for the protogalactic `Aurora' population, finding $5 \times 10^8\,M_\odot$. 
This population was also studied by \cite{Belokurov2022a} and \cite{Conroy2022}, who found that although this ancient `\textit{in-situ}' population is predominantly isotropic, it exhibits a small amount of net prograde rotation.
These results were corroborated in the high-resolution FIRE simulations by \cite{Horta2023}, who found that protogalactic populations in all their MW analogs show weak net rotation aligned with the present-day disk. 

Figure~\ref{fig:519311_protogalaxy} shows the protogalactic stellar population assembled by $z = 6$ in TNG50 halo \#519311. 
The top panel shows the circularity distribution at $z = 6$ (red) and -- for the same stars -- at $z = 0$ (blue). 
Again, these circularities are computed relative to the net angular momentum of each snapshot, not relative to the orientation of the present-day disk. 
Although the kinematics of these protogalactic stars were completely disordered and isotropic in the pre-disk era, they have acquired a prominent tail of prograde orbits by the present. 
Therefore, the protogalaxy itself experiences `spin-up' in the true sense of the word, with protogalactic stars being perturbed by the growing disk, bar, or merger around it to preferentially have mildly prograde orbits \citep[e.g.,][]{Dillamore2022}. 
The bottom panel of Figure~\ref{fig:519311_protogalaxy} shows the present-day edge-on distribution of protogalactic ($z > 6$) stars. 
The distribution is visibly oblate, with an ellipticity $\epsilon \approx 0.7$ at $z = 0$, as compared to $\epsilon \approx 0.3$ at $z = 6$, matching results from FIRE \citep{Horta2023}. 

\subsection{The High-$\alpha$ Disk and In-Situ Halo:\\ Remnants of the Old Disk}\label{sec:tng:olddisk}

\begin{figure}
    \centering
    \includegraphics[width=\columnwidth]{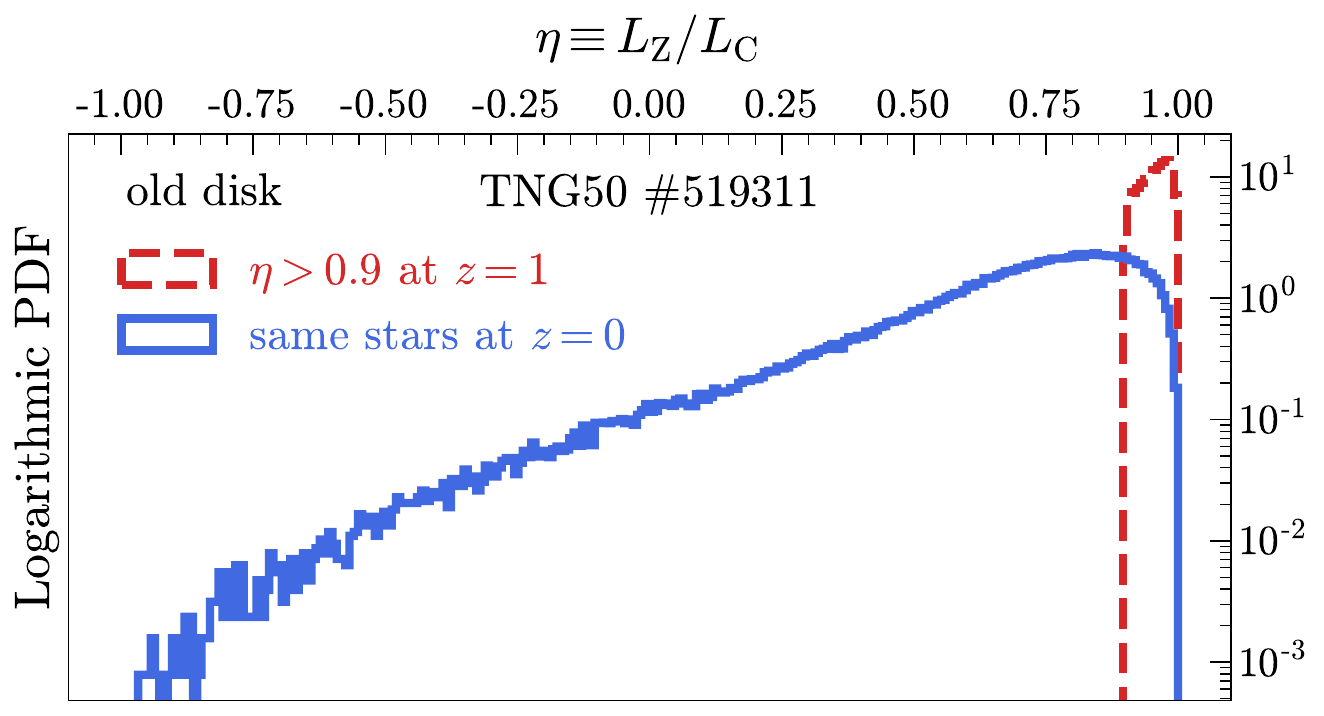}
    \includegraphics[width=0.93\columnwidth]{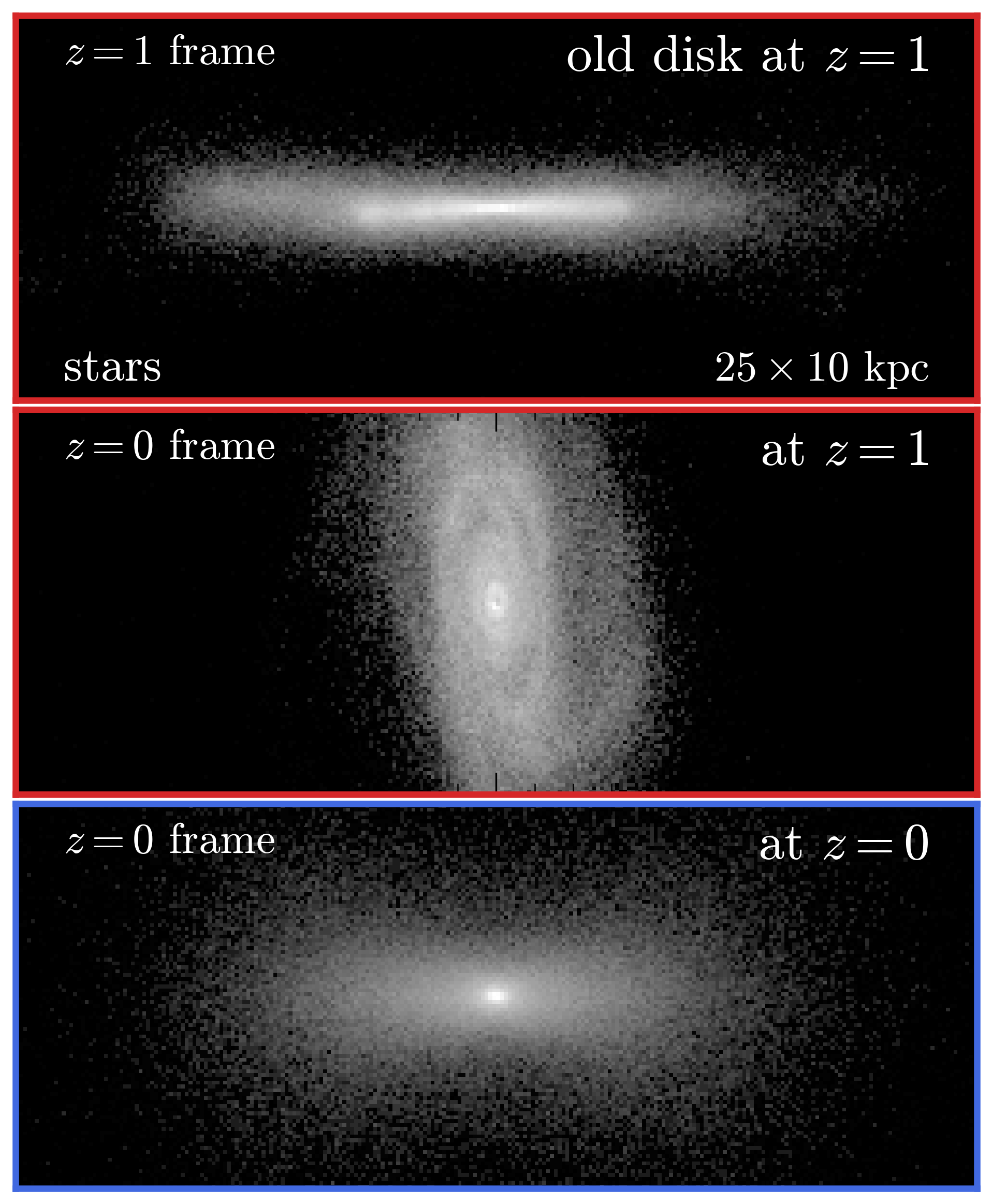}
    \caption{Top: Circularity distribution for old disk stars selected to have $\eta > 0.9$ at $z = 1$ (red), and the distribution for those same stars at the present day (blue). Note that the distribution is logarithmic, so the present day distribution is still disky, albeit with a long tail towards isotropic and retrograde orbits. Bottom: The spatial distribution of the old disk stars at $z = 1$ (red borders) and at $z = 0$ (blue borders). The first image is in an edge-on frame relative to the $z = 1$ galaxy, and the other two are edge-on relative to the $z = 0$ galaxy.}
    \label{fig:519311_olddisk}
\end{figure}

It is apparent from Figures~\ref{fig:portrait} and \ref{fig:wonky} that the Milky Way's high-$\alpha$ disk is kinematically hotter than the younger low-$\alpha$ disk. 
Over time, secular orbital interactions and mergers are expected to vertically `heat' stars to kinematically hotter orbits \citep[e.g.,][]{Quinn1993, Font2001, Kazantzidis2008, Villalobos2008}. 
It therefore remains an open question to what extent stars in the high-$\alpha$ disk were heated after formation, as opposed to, say, being born in more turbulent gas at early times. 
This question was extensively explored in the high-resolution FIRE simulations by \cite{Yu2023}, who found that to first order, stars remain on orbits similar to their birth orbits. 
Mergers and heating processes only play a secondary role in the 12 MW analogs analyzed by \cite{Yu2023}, although those particular galaxies have relatively quiescent merging histories over the past 10\,Gyr. 

Figure~\ref{fig:519311_olddisk} explores the evolution of the old disk in the TNG50 halo  \#519311.  
We identify the `old disk' as all stars on nearly circular orbits ($\eta > 0.9$) that have formed by $z = 1$, just prior to the last major merger (red dashed histogram in the top panel). 
Around half of the $z = 1$ stellar mass of the galaxy is in this highly circular disk component.
As the images in the lower panels illustrate, these stars form a coherent and relatively thin disk in the $z = 1$ `edge-on' frame. 
However, in the $z = 0$ frame---in which the present-day disk would be edge-on---this old disk is nearly face-on (see also Figure~\ref{fig:eta_evol}). 

Moving forward to the present day, these old disk stars have spread out over a large range of orbital circularities, extending all the way down to the retrograde regime (blue histogram in the top panel of Figure~\ref{fig:519311_olddisk}). 
The circularity distribution is reminiscent of the observed present-day `old disk' in the Milky Way (Figures~\ref{fig:etapdf} and \ref{fig:portrait}). 
The spatial distribution resembles a thick disk, ensconced in an oblate halo of scattered stars.
The old disk constitutes $\approx 10\%$ of the present-day stellar mass of this galaxy, comparable to the mass fraction of the MW's `thick' disk \citep{Juric2008, Bland-Hawthorn2016}. 

This illustrative example demonstrates how a present-day thick disk and \textit{in-situ} halo population can be formed by a coherent and thin old disk that is initially misaligned, and significantly torqued by the arrival of a major merger \citep[see also][]{Renaud2021b,Meng2021, Dillamore2022, Dodge2023}. 
Given observational evidence that the Milky Way spun up its old disk remarkably early, and that it underwent a major merger soon afterwards, this kind of disk tilting is a plausible formation channel for the thick disk in our own Galaxy \citep{Bonaca2017, Bonaca2020, Belokurov2020b, Belokurov2022a, Semenov2023}.

\section{Discussion}\label{sec:discuss}

In the beginning, the nascent Milky Way existed as a chaotic and disordered `protogalaxy', constantly bombarded by mergers that built up its most ancient stars \citep[e.g.,][]{Zolotov2009, Tumlinson2010, Pillepich2015, El-Badry2018a, Belokurov2022a, Semenov2023, Horta2023}. 
\cite{Rix2022} used an all-sky sample of giants with \textit{Gaia} low-resolution spectra to reveal this metal-poor population residing in the heart of the MW \citep[building on a rich legacy of observational efforts, e.g.,][]{GarciaPerez2013, Ness2013, Schlaufman2014, Casey2015, Ness2015, Koch2016, GarciaPerez2018, Lucey2019, Arentsen2020a, Arentsen2020b, Reggiani2020, Arentsen2021, Lucey2021, Lucey2022}. 
This protogalactic population is prominent in Figure~\ref{fig:allsky}, and Figure~\ref{fig:etapdf} illustrates how protogalactic stars are net isotropic, albeit with a tail towards prograde orbits. 
$\S$~\ref{sec:tng:protogal} and Figure~\ref{fig:519311_protogalaxy} demonstrate that a protogalactic population in the TNG50 simulations can acquire such a prograde tail after birth, even if it was initially perfectly isotropic \citep[see][]{McCluskey2023, Horta2023}. 

What followed was the most drastic phase transition in our Galaxy's history, the birth of the Galactic disk. 
The ancient disk coincides with the present day `high-$\alpha$' (also, hot or thick) disk, although there is an imperfect and imprecise correspondence between these various classifications. 
The high-$\alpha$ disk could be `thick' due to a combination of thicker birth orbits (e.g., due to a more turbulent ISM at high redshift), and subsequent heating by gravitational interactions, with the relative importance of these mechanisms being an active area of research \citep[][]{ForsterSchreiber2009, Dekel2009b, Bird2013, Stott2016, Bird2021disk, VanDonkelaar2022, Yu2023, McCluskey2023, HamiltonCampos2023}. 
Here differences between the adopted physics across simulations also become relevant.
For example, galaxies in the FIRE simulations experience stronger feedback and more `bursty' star formation than TNG \citep[e.g.,][]{Gurvich2023}. 
The close MW analog that we find in TNG50 (\#519311) forms a kinematically-ordered albeit misaligned disk prior to its last major merger, which is subsequently torqued and `thickened' by the merger (Figures~\ref{fig:eta_evol} and \ref{fig:519311_olddisk}).
Early results from the first year of \textit{JWST} observations have revealed a large population of disks at $z \gtrsim 3-6$, supporting the idea that well-ordered disks can assemble in the early Universe \citep[][]{Ferreira2022, Robertson2023, deGraaff2023, Jacobs2023}. 
 
Around $z \approx 1$, star formation from $\alpha$-enhanced birth material came to an end, yielding to the low-$\alpha$ (or `thin') disk. 
The precise timing of these transitions is uncertain and likely overlapping, with the characteristic observed transition time being around $\approx 8$~Gyr ago \citep{Bonaca2020, Xiang2022}. 
This transition timescale coincides with a `reset' in the metallicity of the disk from \feh{}~$\approx 0$ to \feh{}~$\approx -0.8$ at roughly constant \afe{}~$\approx 0.1$, creating the `bimodal' abundance distribution of the disks (see Figure~10 in \citealt{Haywood2013}, and also \citealt{Bensby2014, Hayden2015, Martig2016, Nissen2020, Conroy2022}). 
Although it remains to be seen how common such a chemical bimodality is in observed Milky-Way-like galaxies \citep[e.g.,][]{Nidever2023, vandeSande2023}, \textit{spatially} distinct disks (thick and thin) are known to be ubiquitous, with the prevalence of old thick disks suggesting a merger-related origin \citep[e.g.,][]{Dalcanton2002}. 

The Milky Way's age pattern in the [Fe/H] -- [$\alpha$/Fe] abundance space is suggestive of a `zig-zag' evolution, with the metallicity `resetting' from the solar-metallicity high-$\alpha$ disk to the metal-poor low-$\alpha$ disk, at near-constant \afe{} \citep{Haywood2013, Bensby2014, Conroy2022}. 
Indeed, recent high-resolution simulations of MW analogs have demonstrated how a major gas-rich merger can produce the chemical bimodality and age trend seen in the Milky Way \citep{Buck2020, Buck2020a, Agertz2021, Renaud2021a, Renaud2021b}. 
However, it has been challenging to associate the  GSE merger definitively with the bimodality of the disks, primarily to the lack of precise ages. 
Different studies have arrived at disparate conclusions about the temporal ordering of the truncation of the high-$\alpha$ disk, arrival of the GSE merger, and birth of the low-$\alpha$ disk \citep{Gallart2019, Bonaca2020, Xiang2022}. 
Improvements in the precision and accuracy of stellar ages will be vital to produce a more fine-grained picture of the early Milky Way. 

\cite{Myeong2022} find evidence for a new eccentric component of the MW halo, which they name Eos. 
Eos stars lie broadly in the low-$\alpha$ sequence around \feh{}~$\approx -1.0$.
At low metallicities, Eos chemically resembles the metal-rich accreted stars from the GSE merger, whereas at higher metallicity Eos connects with the low-$\alpha$ disk.
This population therefore exhibits features that we would expect of the earliest low-$\alpha$ disk stars born soon after the GSE merger, immediately following the metallicity `reset' of the disk. 
This possibility warrants further investigation, chiefly by age-dating Eos stars. 
Furthermore, it remains to be seen whether detailed hydrodynamic simulations of the MW-GSE merger can reproduce such a population in the outer disk, and whether the kinematic properties of Eos match these predictions \citep[see][]{Orkney2022, Buck2023, Rey2023}. 

Regardless of whether a major merger is required to produce the chemical bimodality of the MW disk, it is now quite certain that the MW did undergo such a merger at $z \approx 2$.
One of the most significant results from the \textit{Gaia} era was the confirmation that the majority of the MW's stellar halo was likely deposited by a single major merger, the \textit{Gaia}-Sausage Enceladus \citep[GSE, e.g.,][]{Belokurov2018a, Helmi2018, Haywood2018, Gallart2019, Mackereth2019, Naidu2020, Naidu2021, Myeong2022, Chandra2023}. 
\cite{Bonaca2020} and \cite{Giribaldi2023} find that star formation in GSE truncated $9.5 - 10$~Gyr ago, preceding the formation of the low-$\alpha$ disk by $\approx 2$~Gyr.
This merger could also have dynamically perturbed the ancient disk, creating the observed population of stars on halo-like orbits with chemistry that is indistinguishable from the high-$\alpha$ disk \citep[the in-situ halo or `splash', ][see Figure~\ref{fig:splash}]{Bonaca2020, Belokurov2020b}. 
It is tempting to imagine a unified picture in which the GSE merger mediates the transition between the high-$\alpha$ and low-$\alpha$ disks, providing the pristine gas required to create a chemical bimodality, and dynamically creating the in-situ halo in the process. 
Such a scenario is thus far supported---but not demanded---by current MW data. 

\section{Conclusions}\label{sec:conclude}

We have assembled a dataset of 10 million red giant stars with 6D kinematics, metallicities, and $\alpha$-abundances using data from \textit{Gaia} DR3 (Figures~\ref{fig:allsky}--\ref{fig:etapdf}), and have used it to draw a comprehensive empirical picture of our Milky Way's formation history. 
This picture builds on a rich legacy of prior work in Galactic archaeology, now synthesized with all-sky Gaia data:

\begin{enumerate}

    \item The angular momentum evolution of the Milky Way exhibits three distinct phases as a function of metallicity: the disordered protogalaxy ([Fe/H] $\lesssim -1.2$), the now-hot old disk ($-1.0 \lesssim \text{[Fe/H]}\lesssim -0.7$), and the cold young disk ($\text{[Fe/H]}\gtrsim -0.5$). 
    The transitions between these phases are sharp ($\sim$0.2\,dex in [Fe/H]), and can be described as `spinup' and `cooldown' respectively.  
    The hot old disk and cold young disk are distinctly separable in \afe{}-\feh{} abundance space, mapping onto the high-$\alpha$ and low-$\alpha$ disks respectively (Figure~\ref{fig:portrait}).

    \item The observed protogalaxy is nearly isotropic and mostly confined to the inner heart of the Galaxy, but has mild net rotation even at rather low metallicities (Figures~\ref{fig:etapdf} and \ref{fig:portrait}).
    More detailed individual chemical abundances are required to perform a precise excision of the accreted halo population from the bona-fide protogalaxy in the inner Galaxy. 
    
    \item Since the old, high-$\alpha$ disk is known to have a relatively monotonic age-metallicity relation, we construct a time-ordered portrait of the birth of the Milky Way's ancient disk (Figure~\ref{fig:portrait}, bottom-left).
    After spinning up from the protogalaxy, the high-$\alpha$ disk exhibits a smooth evolution toward increasingly circular orbits throughout its metallicity range $-1.0 \lesssim$~\feh{}~$\lesssim 0.0$ (Figure~\ref{fig:wonky}). 
    However, at \feh{}~$\lesssim -0.5$, high-$\alpha$ stars also occupy a vast swath of orbits extending all the way to the retrograde regime, forming an \textit{in-situ} halo population (Figure~\ref{fig:splash}). 

    \item The young, low-$\alpha$ disk occupies highly circular orbits across all metallicities and exhibits a well-established radial metallicity gradient. 
    The effects of this gradient are best visualized in the bottom-left panels of Figure~\ref{fig:wonky}, which illustrate the strikingly different angular momentum profiles of the high- and low-$\alpha$ disks as a function of metallicity. 
    \setcounter{sum}{\value{enumi}}
\end{enumerate}

In $\S$\ref{sec:tng}, we interpret these observed trends using MW analogs selected from the IllustrisTNG50 cosmological simulations. 
We emphasize the importance of finding \textit{evolutionary} analogs to the Milky Way, instead of only structural analogs at the present day. 
Our key findings are as follows: 

\begin{enumerate}
    \setcounter{enumi}{\value{sum}}
    
    \item \citetalias{Semenov2023} find that out of 61 MW-like galaxies in TNG50, only 10 exhibit an `early spinup' in rotational velocities that match MW data. 
    We find that although most of these early spinup analogs exhibit a three-stage evolution in the circularity--\emph{age} space, only one (halo \#519311) matches our observations of a three-phase behavior in the circularity--\emph{metallicity} plane (Figure~\ref{fig:tng_portrait}). 
    This galaxy is notable for a particularly gas-rich major merger $\approx 7$~Gyr ago---reminiscent of (but probably later than) the MW's own merger with GSE---and a relatively high metallicity of the young cold disk (Section~\ref{sec:tng:analog}, Figure~\ref{fig:tng_mah}). 

    \item The time-resolved evolution of this simulated MW analog illustrates that it underwent an early spinup $\approx 11$~Gyr ago, and had a kinematically ordered disk well in place by $\approx 8$~Gyr ago. 
    The subsequent gas-rich merger tilted and heated the ancient disk, and a new kinematically cold disk subsequently formed, leading to the `three-phase' kinematic signature (Figure~\ref{fig:eta_evol}).

    \item The simulated MW analog exemplifies that protogalactic populations can flatten and acquire net rotation due to external forces from the surrounding disk, even if the protogalaxy was originally completely isotropic (Figure~\ref{fig:519311_protogalaxy}). 
    It also demonstrates how an initially misaligned, kinematically cold old disk can be tilted and heated by a major merger, creating both a spatially thick disk and in-situ halo component (Figure~\ref{fig:519311_olddisk}). 

    \item Overall, the MW analog from TNG50 demonstrates how a single decisive major merger striking an early-formed galactic disk can reproduce various kinematic properties of the Milky Way. Such a merger-driven formation scenario for the disks is therefore strongly supported---but not yet demanded---by the observed data. 

\end{enumerate}

Our work demonstrates that the three dominant evolutionary stages of the Milky Way---the protogalaxy, old disk, and young disk---are inextricably linked by the three physical processes of spinup, merger, and cooldown. 
Whereas there is now significant evidence from the halo that our Galaxy underwent a major merger $\approx 10$~Gyr ago, here we show that this merger could have played a decisive role in transforming the Galactic disks as well. 
Although several pieces of the puzzle are still missing---chiefly precise stellar ages that could support the ordering and causal links between phases---this simple picture does a remarkable job reproducing the overall kinematic structure of the Milky Way. 
A concert of future surveys and simulations will hopefully further illuminate a plausible biography of the Galaxy we call home. 

\newpage

\begin{acknowledgments}

VC gratefully acknowledges a Peirce Fellowship from Harvard University.  
It's a pleasure to thank  
Angus Beane,
Jonathan Bird, 
James Bullock,
Kareem El-Badry,
Michael Hayden,
David W. Hogg,
Phil Hopkins,
Adrian Price-Whelan,
Sandro Tacchella,
and Turner Woody
for insightful conversations and feedback. 
This work benefited from discussions at the 2022 Disk Formation Workshop at UC Irvine, the 2023 Gaia XPloration workshop at the University of Cambridge, and the 2023 Galaxy Formation $\times$ Theory Workshop at Biosphere 2. 
Support for V.S. was provided by Harvard University through the Institute for Theory and Computation Fellowship. 
RPN acknowledges support for this work provided by NASA through the NASA Hubble Fellowship grant HST-HF2-51515.001-A awarded by the Space Telescope Science Institute, which is operated by the Association of Universities for Research in Astronomy, Incorporated, under NASA contract NAS5-26555.  
LH acknowledges support by the Simons Collaboration on “Learning the Universe”.

This work has made use of data from the European Space Agency (ESA) mission {\it Gaia} (\url{https://www.cosmos.esa.int/gaia}), processed by the {\it Gaia} Data Processing and Analysis Consortium (DPAC, \url{https://www.cosmos.esa.int/web/gaia/dpac/consortium}). Funding for the DPAC has been provided by national institutions, in particular the institutions participating in the {\it Gaia} Multilateral Agreement. 
The computations in this paper were run on the FASRC Cannon cluster supported by the FAS Division of Science Research Computing Group at Harvard University.
This research has made extensive use of NASA's Astrophysics Data System Bibliographic Services.

\end{acknowledgments}

\software{\texttt{numpy} \citep{Harris2020}, 
\texttt{scipy} \citep{Virtanen2020}, 
\texttt{matplotlib} \citep{Hunter2007}, 
\texttt{gala} \citep{gala,adrian_price_whelan_2020_4159870}
}

\facilities{Gaia, WISE}

\bibliography{library,bib}
\bibliographystyle{aasjournal}

\end{document}